\documentclass[aps,prd,onecolumn,amsmath,amssymbepsfig,showpacs,nofootinbib]{revtex4}

\usepackage{graphicx}
\usepackage{bm}
\usepackage{epsfig}
\usepackage{pifont} 
\usepackage{yfonts}
\usepackage{dcolumn}
\usepackage{natbib}

\fontencoding{T1}

\def\be{\begin{equation}}
\def\ee{\end{equation}}
\def\beq{\begin{equation}}
\def\eeq{\end{equation}}

\def\ba{\begin{eqnarray}}
\def\ea{\end{eqnarray}}

\def\nn{\nonumber}

\newcommand{\mb}[1]{\mathbf{#1}}

\newcommand{\fnl}{f_{\mathrm{NL}}}
\newcommand{\gnl}{g_{\mathrm{NL}}}
\newcommand{\tnl}{\tau_{\mathrm{NL}}}

\begin{document}

\title{Structure formation from non-Gaussian initial conditions: \\ 
multivariate biasing, statistics, and comparison with N-body simulations}

\author {Tommaso~Giannantonio}
\email{giannantATastroDOTuni-bonnDOTde,  porcianiATastroDOTuni-bonnDOTde}
\affiliation{Argelander--Institut f\"ur Astronomie der Universit\"at Bonn, Auf dem H\"ugel 71, D-53121 Bonn, Germany}

\author {Cristiano~Porciani}
\affiliation{Argelander--Institut f\"ur Astronomie der Universit\"at Bonn, Auf dem H\"ugel 71, D-53121 Bonn, Germany}

\begin{abstract} 
We study structure formation in the presence of primordial non-Gaussianity of the local type
with parameters $\fnl$ and $\gnl$. We show that the distribution of dark-matter halos is naturally described by a multivariate bias scheme where the halo overdensity depends
not only on the underlying matter density fluctuation $\delta$ but also on the Gaussian part of
the primordial gravitational potential
$\varphi$.
This corresponds to a non-local bias scheme in terms of $\delta$ only.
We derive the coefficients of the bias expansion as a function of the halo
mass by applying the peak-background split to common parameterizations
for the halo mass function in the non-Gaussian scenario.
We then compute the halo power spectrum and halo-matter cross
spectrum in the framework of Eulerian perturbation theory up to third order.
Comparing our results against N-body simulations, we find that
our model accurately describes the numerical data for 
wavenumbers $k\leq 0.1-0.3 \ h$ Mpc$^{-1}$ depending on redshift and
halo mass.
In our multivariate approach, perturbations in the halo counts
trace $\varphi$ on large scales and this explains why the halo and matter
power spectra show different asymptotic trends for $k\to 0$.
This strongly scale-dependent bias 
originates from terms at leading order in our expansion.
This is different from 
what happens using the standard univariate local bias
where the scale-dependent terms come from badly behaved higher-order 
corrections. 
On the other hand, our biasing scheme
reduces to the usual local bias on smaller scales where $|\varphi |$ is typically much
smaller than the density perturbations. 
We finally discuss the halo bispectrum in the context of multivariate
biasing and show that, due to its strong scale and
shape dependence, it is a powerful tool for the detection of primordial
non-Gaussianity from future galaxy surveys. 
\end{abstract}


\maketitle

\label{firstpage}

\section{Introduction}
Measurements of the cosmic microwave background (CMB) anisotropies have confirmed the hypothesis that the present inhomogeneities in the matter density were seeded by small fluctuations at primordial times \citep{Komatsu:2008hk}. Such perturbations, which are expected to be created as quantum vacuum fluctuations,
have been generally modeled with 
the simple statistical assumption of being 
a Gaussian random field with nearly scale invariant power spectrum \cite{Bardeen:1985tr}.

The inflationary mechanism is often invoked to describe the early 
universe, but the details
remain 
debated 
\cite {Lyth:1998xn}.
In the simplest single-field, slow-roll model, small curvature (adiabatic) perturbations are generated with a nearly Gaussian distribution \cite{Maldacena:2002vr, Acquaviva:2002ud}. However in other models such as the curvaton scenario  \cite{Lyth:2001nq, Linde:1996gt, Lyth:2002my} some additional fields would decay 
at later times,
producing larger non-Gaussianity \cite{Sasaki:2006kq,  Malik:2006pm, Vernizzi:2006ve}; cyclic or ekpyrotic universes without inflation could also produce large non-Gaussianities during their contracting phase \cite{Khoury:2001wf,Creminelli:2007aq}.
 Furthermore, multi-field models can in general produce isocurvature modes of the perturbations \cite{Polarski:1994rz}.
 See \cite{Bartolo:2004if} for a
 review and \cite{Komatsu:2009kd} for 
recent updates and future prospects.

The first observable predictions from inflation --- flatness and the near scale invariance of the power spectrum of the perturbations --- have been under scrutiny for some time from observations of both the large scale structure (LSS) \cite{Tegmark:2006az} and the CMB \cite{Komatsu:2008hk}. Fairly strict constraints on the adiabaticity of the primordial perturbations also exist \cite{Valiviita:2009bp}, while their Gaussian distribution has only recently become testable.

Although other possibilities exist (see \cite{Fergusson:2008ra} for a review),
many models produce primordial non-Gaussianity of the \emph{local type} where
the Bardeen's potential $\Phi$ can be expressed in terms of an auxiliary
Gaussian potential $ \varphi$ as
\be \label{eq:basichi}
\Phi (\mathbf{x}) = \varphi(\mathbf{x}) + \sum_{j=2}^{\infty} Q_{\mathrm{NL}j} \left[ \varphi^j(\mathbf{x}) - \langle \varphi^j(\mathbf{x}) \rangle \right],
\ee
where the series will be in practice truncated at some finite order $N$,
and the odd momenta of the Gaussian potential $\varphi$ vanish by definition.
The first parameter $Q_{\mathrm{NL}2}$, usually dubbed 
$\fnl$, quantifies the leading-order departure from purely Gaussian initial
conditions through the irreducible three-point function or the bispectrum of
the potential.
While standard inflation forecasts a slow-roll suppressed, primordial
$|\fnl| \ll 1$, subsequent evolutionary processes are expected to increase the
amount of non-Gaussianity up to $|\fnl|\sim 1$ \cite{Bartolo:2003gh}.
On the other hand, 
more complex models can produce $ |\fnl| \gg 1 $,
although the actual predicted values vary.
The second parameter $Q_{\mathrm{NL}3}$, generally 
called $\gnl$,
quantifies the next higher order contribution and is related to the
irreducible four-point function or the trispectrum of the potential. Since
$\varphi \sim 10^{-5} $, this contribution can be 
important only if $\gnl$ is big,
 $ \gnl \gtrsim \fnl^2$. 
This is
plausible in the interactive curvaton model \cite{Enqvist:2008gk,Assadullahi:2007uw,Huang:2008zj} and in other multi-field 
 scenarios \cite{Byrnes:2009qy,Huang:2009vk}.

The traditional method to constrain primordial non-Gaussianity has been the 
three-point statistics of CMB anisotropies. The current limits from
WMAP are $-9 < \fnl < 111 $ at the $95\%$ c.l. \cite{Komatsu:2008hk}. Different analyses of the same data found $ -178 < \fnl < 64 $ using Minkowski functionals \cite{Komatsu:2008hk}, $-4 < \fnl < 80$ using an optimized estimator \cite{Smith:2009jr}, and $ -18  < \fnl <  80 $ from wavelet decomposition \cite{Curto:2009pv},
while a detection ($ 27 < \fnl < 247 $) was claimed by \cite{Yadav:2007yy}. 
Constraints on $\gnl$ are $-5.6 \cdot 10^5 < \gnl < 6.4 \cdot 10^5$ \cite{Vielva:2009jz}. 
The Planck satellite
is expected
to reduce the uncertainty to $ \sigma (\fnl) \sim 5 $ \cite{Komatsu:2001rj}. 
This result will be nearly cosmic-variance limited, 
and further significant 
improvements from CMB studies will be 
difficult to achieve. 
This reason, together with the desire of having independent results, affected by different systematics, provided the motivation to study the detectability of primordial non-Gaussianity from the LSS.
In this case,
however, the non-linear growth of density perturbations can superimpose a new non-Gaussian signal onto the primordial one \cite{Scoccimarro:2003wn}, which may be difficult to retrieve.
Determining the mass distribution of galaxy clusters at low and high redshift
provides a way to circumvent this problem \cite{Matarrese:2000iz,LoVerde:2007ri}. 
However, due to the low-number statistics, these methods have been so far less
successful than the CMB. 
Upcoming surveys such as PanSTARRS, DES, LSST, ADEPT,
EUCLID, JDEM or eROSITA, WFXT and SPT are expected to substantially improve the situation.

A new technique, based on linear perturbation theory, has been recently introduced by \cite{Dalal:2007cu} (Dal07 in
the following). 
These authors showed that
local non-Gaussianity breaks the independence of small and large scales density fluctuations. As a consequence, the clustering of dark matter halos is altered, 
becoming 
enhanced on large scales for a positive $\fnl$.
An analytical derivation of the corresponding scale-dependent bias has been
also presented by \cite{Matarrese:2008nc, Slosar:2008hx,Afshordi:2008ru,Valageas:2009vn},
together with some observational constraints on $\fnl$ from existing data of
the clustering of galaxies and their correlation with the CMB anisotropies. Using luminous
red galaxies and quasars from the 
SDSS, 
\citet{Slosar:2008hx} obtained $ -29 < \fnl < 69 $, 
 competitive
with the CMB results.
The first constraints on $\gnl$ 
 from the
LSS give $ -3.5 \cdot 10^5 < \gnl < 8.2 \cdot 10^5 $ \cite{Desjacques:2009jb},
\emph{assuming $\fnl = 0$}.

N-body simulations
show only approximate agreement with the model by Dal07
\cite {Desjacques:2008vf,
  Pillepich:2008ka, Grossi:2009an}. 
In particular, the power spectrum of dark-matter halos seems to scale
with the wavenumber and the $\fnl$ parameter in a different way than predicted.
This discrepancy provides the main
motivation for this paper where we 
study the effect of non-Gaussian initial conditions on the clustering of halos
in the weakly non-linear regime of perturbation growth.
Applying the peak-background split technique, we show that the halo distribution on large scales is naturally
described by a \emph{bivariate local biasing scheme}, where the halo overdensity is
expanded in a Taylor series of both the matter perturbations $\delta$ and the primordial Gaussian
potential $\varphi$. Since $\varphi$ and $\delta$ are related by the Poisson equation, this can be equivalently seen as a non-local description in terms of $\delta$ only.
This reduces to the usual univariate \emph{local} bias (where halo overdensities are expanded in
series of $\delta$ only) for Gaussian initial conditions and, in
general, on small scales.
Using standard (Eulerian) perturbation theory (SPT) up to third order to
account for the non-linear growth of density fluctuations, we show that our
new biasing scheme leads to the presence of several new terms in the halo
power spectrum and bispectrum.
We show that our results reduce to the usual Gaussian solution in the limit $\fnl
\rightarrow 0$, 
and to the results by Dal07 (revised as in \cite{Slosar:2008hx,Desjacques:2008vf}) if we only consider the leading-order terms.
Finally, we compare our theory with the N-body simulations by \cite{Pillepich:2008ka}
(hereafter PPH08) and find that our model can explain the
numerical results to a much greater accuracy than both linear and univariate local theories.

Our paper differs from the recent work by 
\cite{Taruya:2008pg, Sefusatti:2009qh, Sefusatti:2007ih, Jeong:2009vd}
which is based on the assumption 
that fluctuations in the halo number counts only depend on the local mass density.
By ignoring the expansion in the potential $\varphi$
 and only considering the dependence on the matter density
perturbations $\delta$, 
one would obtain different results which do not reduce to the model by Dal07
to leading order and do not match the simulations as well;
 in this case 
higher-order terms in the halo power spectrum grow bigger 
than the first-order contribution
on large scales, thus casting 
doubts on the validity of the perturbative 
expansion.
Our approach is also different from the work by \cite{McDonald:2008sc}
because we use SPT without
 applying any 
 renormalization technique. 
Renormalizing the bias 
removes any undesired dependence on 
the cutoff scale introduced to regularize loop corrections in SPT; 
however, in such a model the bias coefficients cannot be predicted
and should be used as fitting parameters to match observations or simulations.

The plan of this paper is as follows.
In Section \ref {sec:tracers} we summarize the main
biasing schemes which have been proposed to describe the distribution of
different tracers of the LSS.
In Section \ref{sec:massfunc} we introduce some models for the halo mass function
arising from non-Gaussian initial conditions 
and compare them against the N-body results by PPH08. 
We then describe in Section \ref {sec:bias}
how a multivariate bias scheme naturally emerges by applying the
peak-background split technique to compute halo overdensities in the
non-Gaussian case. 
In Section \ref{sec:statistics}
we give a short summary of the statistical properties of non-Gaussian density fields.
After computing 
the halo power spectrum and the halo-matter cross spectrum in our multivariate biasing scheme
in Section \ref{sec:pk}, 
we test our theoretical models against the N-body simulations by PPH08. 
We derive the halo bispectrum 
in Section \ref{sec:bispectrum}, and conclude in Section \ref {sec:concl}.

\section {Tracers of the large-scale structure and biasing } \label{sec:tracers}
The large-scale structure of the Universe can be described in terms of
different tracers: mass, luminosity, galaxy counts. In this
paper we will consider dark-matter halos and mass but our formalism can be
straightforwardly extended to any other tracer.
Let us consider the mass overdensity field $ \delta (\mb x) $ 
and the corresponding density contrast of dark-matter halos in a given mass 
range $\delta_h(\mb x)$.
After smoothing both fields on a relatively large scale $R$, it is reasonable to 
expect that $\delta_h$ is a local function of $\delta$ that can be expanded in a
Taylor series as follows: 
\be
\delta_h(\mb x)=b_0+b_1 \delta(\mb x)+b_2\delta^2(\mb x)/2!+b_3\delta^3(\mb
x)/3!+\dots\;,
\label{ldb}
\ee
where the bias coefficients $b_i$ are in principle scale and mass dependent 
\cite{Fry:1992vr}.
This approximation neglects stochasticity in the $\delta_h$ vs. $\delta$ relation
and is thus dubbed local 
deterministic biasing. Numerical simulations from Gaussian initial conditions 
show that it is accurate on scales of the order of
10 Mpc and larger \cite{2001MNRAS.320..289S}.
In the following sections we will show that Eq.~(\ref{ldb}) does not hold in 
the presence of primordial non-Gaussianity of the local type and 
we will explain how it should be modified.

Note that, in general, 
the bias coefficients in Eq.~(\ref{ldb}) are not independent as the mean halo 
overdensity must vanish and $\delta_h$ must assume the value $-1$ when
$\delta=-1$. 
In order to build a predictive theory, the values for the bias coefficients
should be derived from a model. A common approach is to use the
peak-background split technique \cite {Bardeen:1985tr, Cole:1989vx, Mo:1995cs, Catelan:1997qw}
where the mass perturbations are divided into fine-grained (peak) and 
coarse-grained (background) components. The key idea is to ascribe halo
formation to the collapse of the high-frequency modes, while the large-scale
distribution and motion of these condensations is determined by 
the low-frequency modes.
Starting from a model for the conditional halo mass function (i.e. the
mass function in regions where the background density assumes a specific value),
the peak-background split gives an expression for the halo distribution
in Lagrangian space
(i.e. in the linear density field, $\delta_1$): 
\be
\delta_h^L(\mb q)=b_0^L+b_1^L \delta_1(\mb q)+b_2^L\delta_1^2(\mb
q)/2!+b_3^L\delta_1^3(\mb q)/3!+\dots\;,
\label{llagbias}
\ee
where the bias coefficients $b_i^L$ are obtained from the $i$-th order 
derivatives of the conditional mass function with respect to the background
density contrast (see  Section \ref{sec:bias} for further details).
When the background scale is much larger than the Lagrangian size of the halos,
the Lagrangian bias parameters show very little dependence on the background
scale and the unconditional mass function can be safely used to derive them
\cite{Sheth:1999mn}. We will follow this approach in this paper.

In the absence of large-scale velocity bias, 
the halo density in the evolved Eulerian space is given by
\be \label {eq:eulag4}
1+\delta_h(\mb x)=[1+\delta_h^L(\mb q)][1+\delta(\mb x)]
\ee
where $\mb q$ is the Lagrangian position of the fluid elements that moved
to the Eulerian location $\mb x$ \cite{Catelan:1997qw}.
Note that the conversion between Lagrangian and Eulerian quantities is 
non-local, non-linear and stochastic as it depends on the displacement 
$\mb x-\mb q$ and on both the initial and the evolved fields $\delta_1$ and 
$\delta$. Therefore, the local Lagrangian biasing scheme given in Eq.~(\ref{llagbias})
 will not generally be compatible with Eq.~(\ref{ldb}). \citet{Catelan:2000vn} showed that these
 two bias models give rise to
different shapes of the halo bispectrum that could then be used to
distinguish between them using data from observations or simulations.
A simplified approach is obtained by assuming that the long-wavelength
modes of the density field evolve locally according to the spherical collapse 
model \cite{Mo:1995cs, Mo:1996zb}. In this case, 
Eqs.~(\ref{llagbias}) and (\ref{ldb}) are fully compatible and the
Eulerian bias parameters can be written in terms of the Lagrangian ones
(see Eq.~(\ref{eq:eubi}) in Section \ref{sec:bias}). In particular, $b_1=1+b_1^L$. 
This equation is completely general as it derives from 
mass conservation \cite{Mo:1995cs, Catelan:1997qw}. 
However, the relation between higher-order Eulerian and 
Lagrangian bias parameters depends on the adopted dynamics for the background
density field.
A perturbative calculation of the power spectrum for local Lagrangian biasing 
in the Gaussian scenario has been presented by \cite{Matsubara:2008wx}.
The equivalent result for the local Eulerian bias scheme has been derived
by \cite{Heavens:1998es}. 
In this paper we generalize this latter result to non-Gaussian
initial conditions of the local type 
and also present a model for the bias coefficients as a function of the halo mass.
The derivation of a multivariate bias scheme and the corresponding
calculations of the halo power spectrum and bispectrum constitute our main results.

\section {Halo mass function and primordial non-Gaussianity}  \label{sec:massfunc}

The number density $\mathcal {N}$ of halos of mass $M$ at a redshift $z$ is described by the mass function
\be
n = \frac {d \mathcal {N}} {dM} = f \left(\frac {\delta_c} {\sigma} \right)
\frac {\bar \rho} {M^2} \left| \frac {d \ln \sigma^{-1}} {d \ln (M)}
\right|\;,
\label{eq:massfun}
\ee
where $\delta_c\simeq 1.686$ is the threshold for the linear density contrast
which corresponds to the collapse of spherical perturbations. 
In Eq. (\ref{eq:massfun}),
$\sigma^2$ denotes the variance of the linear density field, calculated as
\be
\sigma^2 (M,z) = \frac{D^2(z)}{2 \pi^2}\int k^2 \, P_0 (k) \, W^2_f (k, M) dk,
\ee
with $P_0(k)$ the linear matter power spectrum at $z=0$, $D(z)$ the linear
growth factor of density fluctuations normalized to unity today, and $W_f(k,M)$ a filter function
with mass resolution $M$. We use a top-hat filter in real space with radius $R_f
= [3 M / (4 \pi \bar \rho)]^{1/3}$, where $\bar \rho$ is the average density of the
Universe. 
The analytical form of the distribution $f(\delta_c / \sigma)$ can be derived from a theoretical
model or by fitting numerical data. We list below some possible choices for this function.

\subsection {Gaussian mass functions}

\begin {figure} 
\begin{center}
\includegraphics[width=0.32\columnwidth,angle=0]{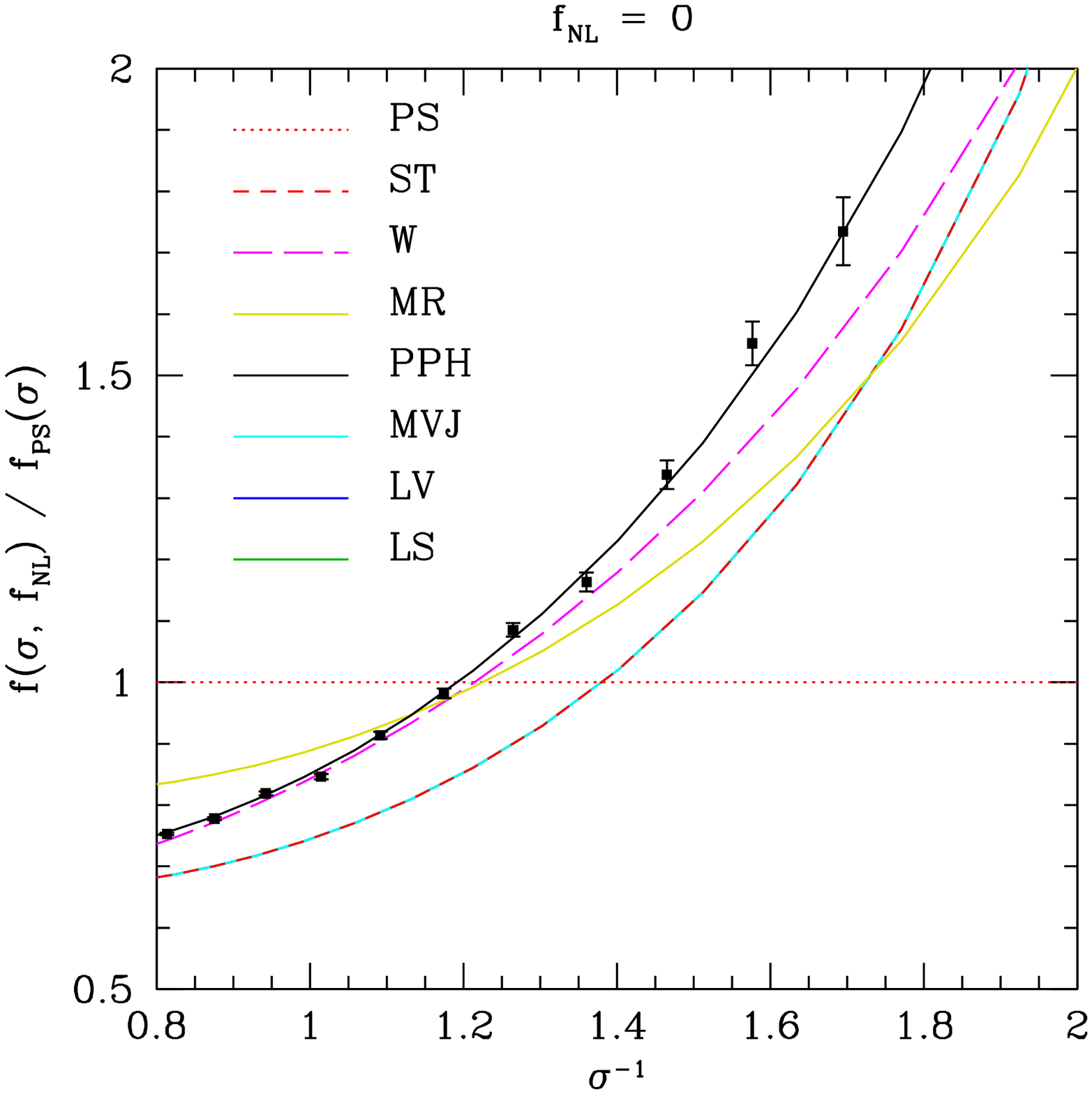} 
\includegraphics[width=0.32\columnwidth,angle=0]{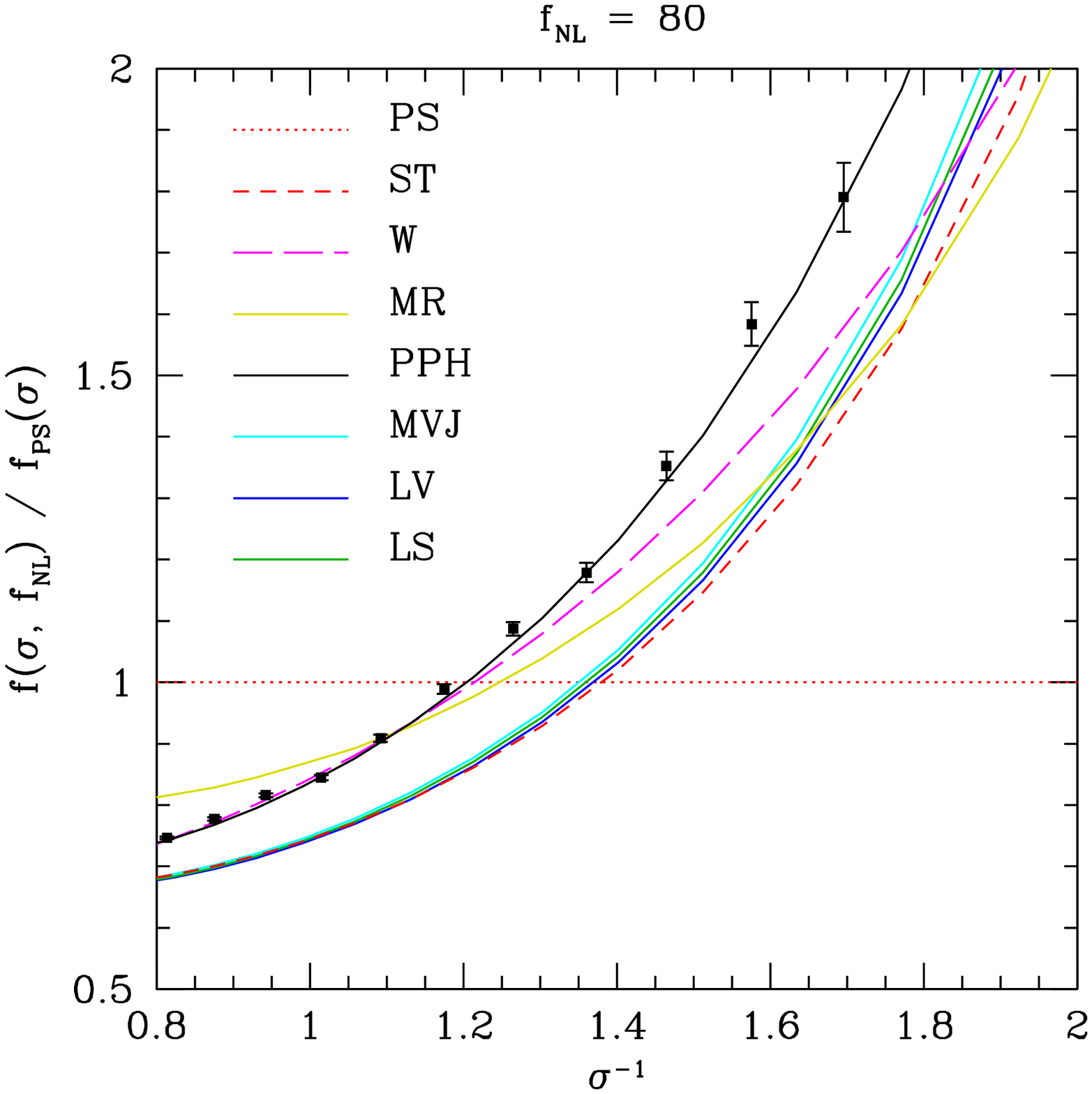} 
\includegraphics[width=0.32\columnwidth,angle=0]{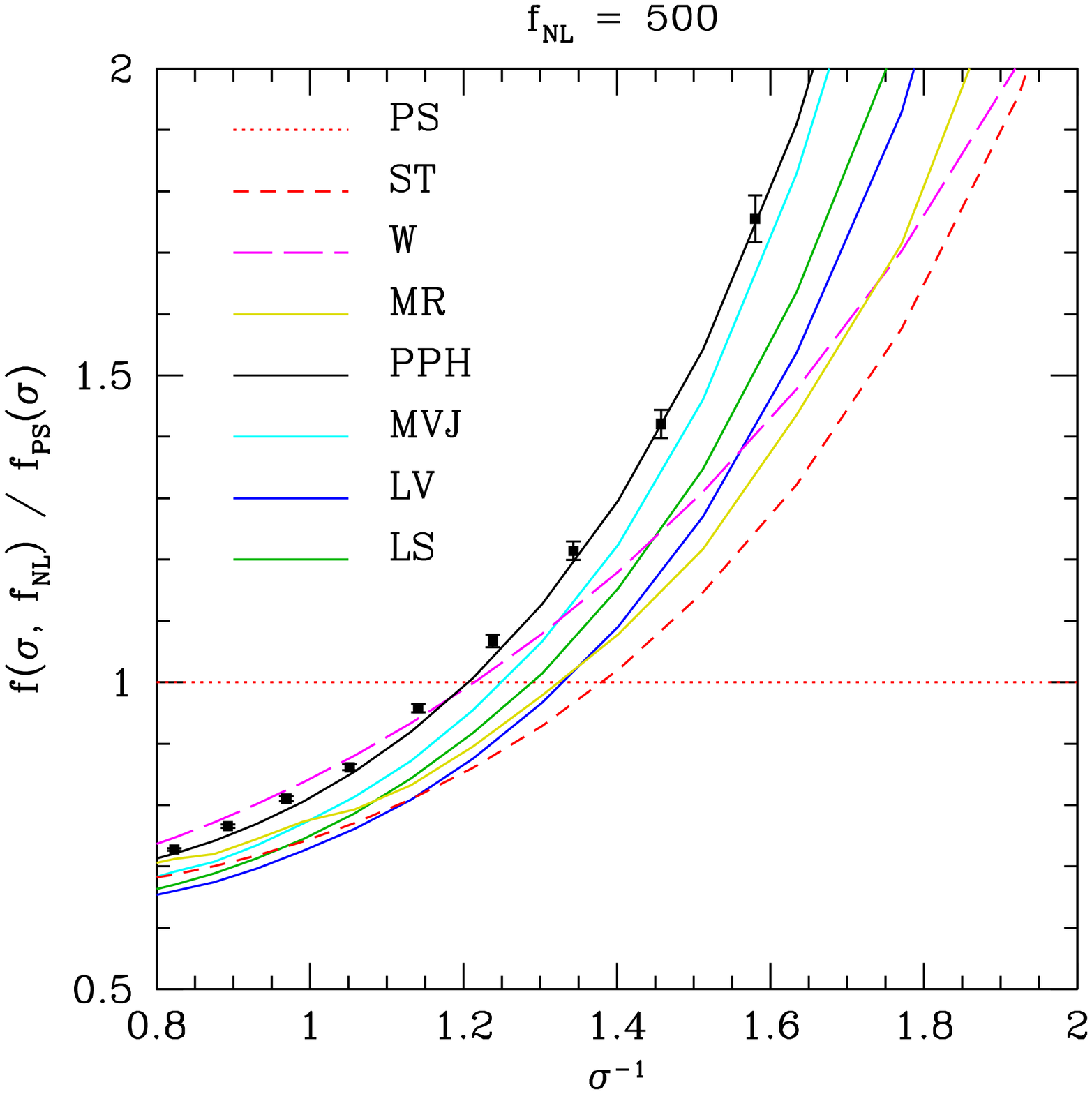}
\end{center}
\caption{Comparison of different models for the halo mass function originating
  from  Gaussian (dashed) and non-Gaussian (solid) initial conditions with the
  N-body data by PPH08. From left to right, the different panels refers to
  $\fnl = 0, 80, 500$. Note that all non-Gaussian models (with the exception  of MR) have been rescaled by the ratio $f_{\mathrm {ST}} / f_{\mathrm {PS}} $, and thus coincide with the ST function in the leftmost panel. }
\label{fig:mf}
\end{figure}

\paragraph* {\bf Press-Schechter (PS)}
For reference we first consider the Press-Schechter theory
\citep{Press:1973iz}, in which the mass function 
deriving from Gaussian initial conditions is given by
\beq
f_{\mathrm {PS}} \left( \frac {\delta_c} {\sigma} \right) = \sqrt{\frac{2}{ \pi}} \frac{\delta_c}{\sigma} e^{-\frac{{\delta_c^2}}{2{\sigma}^2}}.
\eeq
It is well known that this model gives only a rough approximation to numerical
data (see Fig.~\ref{fig:mf}).
The Press-Schechter theory can be improved by introducing extra parameters in
the mass function and fitting them against numerical simulations. 
This approach has been followed e.g. by Jenkins et al. 
\cite{Jenkins:2000bv}, Warren et al. (W) \cite{Warren:2005ey},
 Tinker et al. \cite{Tinker:2008ff}.

\paragraph* {\bf Sheth-Tormen (ST)}

The Press-Schechter theory is based on the spherical collapse model. This can
be improved upon by considering the collapse of ellipsoidal perturbations 
and fitting some new parameters against numerical simulations.
\citep{Sheth:1999mn, Sheth:2001dp}.
The final result, known as the ST mass function, is
\be
f_{\mathrm {ST}} \left( \frac {\delta_c} {\sigma} \right) = A \sqrt {\frac{2 \alpha}{\pi}} \left[ 1 + \left( \alpha \frac{\delta_c^2}{\sigma^2} \right)^{-p} \right]  \frac{\delta_c}{\sigma} e^{-\frac{\alpha {\delta_c^2}}{2{\sigma}^2}},
\ee
where the extra parameters are $\alpha = 0.707$, $p = 0.3$ and $A$ is obtained
by requiring that all the mass is collapsed into halos, which gives $A = 0.322$.

\subsection {Non-Gaussian mass functions}
In the simplest case, local
non-Gaussianity is described by truncating Eq.~(\ref{eq:basichi}) after the
second order term, which corresponds to:
\be \label{eq:basic}
\Phi (\mathbf{q}) = \varphi(\mathbf{q}) + 
\fnl \left[ \varphi^2(\mathbf{q}) - \langle \varphi^2(\mathbf{q}) \rangle \right].
\ee
It can be shown that the halo mass function is very sensitive to the value of 
$\fnl$.
For positive (negative) values of $\fnl$, 
its high-mass tail becomes more (less) prominent
than in the Gaussian case.
To first-order in the non-linearity parameter $\fnl$,
it is possible to account for this effect by considering the skewness 
of the density perturbations, defined as
$S_3 (\sigma) = \langle \delta^3 \rangle/\sigma^4$.

\paragraph*{\bf Matarrese-Verde-Jim\'enez (MVJ)} 

The Press-Schechter theory can be generalized to non-Gaussian initial
conditions by using the saddle point approximation to calculate the
probability for the linear density field to be above $\delta_c$
\citep{Matarrese:2000iz}.
In this case, one obtains:
\be
f_{\mathrm {MVJ}} \left( \frac {\delta_c} {\sigma} \right) = \sqrt{ \frac{2}{\pi}} e^{-\delta_{\star}^2 / (2 \sigma^2)} \left|  \frac {\delta_c^3}{6 \, \sigma \, \delta_{\star} } \, \frac{d S_3(\sigma)}{d \ln \sigma} + \frac {\delta_{\star}}{\sigma} \right|,
\ee
where the new parameter $\delta_\star$ is defined as
$\delta_{\star} \equiv \delta_c  \sqrt{1 - \delta_c
  S_3 (\sigma) / 3} $.
Since the ST model outperforms the PS one in the Gaussian case, it is standard
practice to define a new mass function as $ f_{\mathrm {MVJ}} \to f_{\mathrm {MVJ}}
\cdot f_{\mathrm {ST}} / f_{\mathrm {PS}} $. 
A further modification which has been suggested by \cite{Grossi:2009an} 
to improve the agreement with numerical simulations
is to
correct the collapse 
threshold $\delta_c$ by a factor $\sqrt a = \sqrt {0.8}$ in the expression
 $f_{\mathrm {MVJ}}/f_{\mathrm {PS}}$. In what follows we will adopt both corrections.
A similar result was derived by \cite{Valageas:2009vn} using a different approach.

\paragraph*{\bf LoVerde et al. (LV)} \label {sec:LoV} 
Another way to generalize the PS model is to use the Edgeworth expansion to
approximate the probability distribution function for the linear density contrast
\cite {LoVerde:2007ri}. This gives:
\be
f_{\mathrm {LV}} \left( \frac {\delta_c} {\sigma} \right) = \sqrt{ \frac{2}{\pi}} e^{-\delta_c^2 / (2 \sigma^2)} \left\{  \left[ \frac{\delta_c}{\sigma} + S_3(\sigma) \frac{\sigma}{6} \left( \frac{\delta_c^4}{\sigma^4} - 2 \frac{\delta_c^2}{\sigma^2} - 1  \right)   \right] 
+ \frac{1}{6} \frac {d S_3(\sigma)}{d \ln \sigma} \sigma \left( \frac{\delta_c^2}{\sigma^2} - 1 \right) \right\} \;.
\ee
As for the MVJ case, we will use an effective form of this mass function
expressing the correction to the ST formula:  $ f_{\mathrm {LV}} \to f_{\mathrm {LV}}
\cdot f_{\mathrm {ST}} / f_{\mathrm {PS}} $,
with the further modification of correcting the collapse threshold $\delta_c$
by a factor 
$\sqrt a =  \sqrt {0.8}$ in the ratio $f_{\mathrm {LV}}/f_{\mathrm {PS}}$.

\paragraph*{\bf Maggiore-Riotto (MR)} 
Maggiore \& Riotto \citep {Maggiore:2009rx} computed the halo mass function by
solving the excursion set problem for non-Markovian processes with a
path-integral approach, and found
\be
f_{\mathrm {MR}} \left( \frac {\delta_c} {\sigma} \right) = (1 - \tilde \kappa) \sqrt{ \frac{2}{\pi}} \frac {\sqrt a \delta_c}{\sigma} e^{-a \delta^2_c / (2 \sigma^2)} \left[ 1 + \frac {\sigma^2}{6 \sqrt {a} \delta_c} h_{NG} (\sigma)  \right] 
+ \frac {\tilde \kappa}{\sqrt {2 \pi}} \frac {\sqrt {a} \delta_c}{\sigma} \Gamma \left(0, \frac {a \delta_c^2}{2 \sigma^2} \right).
\ee
Here $\Gamma (0,x)$ is the incomplete Gamma function, the additional
parameters are $a \simeq 0.8$ and $\tilde \kappa = a \kappa$ where $\kappa
\simeq 0.4562 - 0.0040\, R_f$ with $R_f$ the smoothing scale. 
Primordial non-Gaussianity affects the function
\be
h_{NG}(\sigma) = \frac{a^2 \delta_c^4}{\sigma^4} S_3(\sigma) - \frac{a \delta_c^2}{\sigma^2} \left[ 2 S_3(\sigma) + U_3(\sigma) - \frac {d S_3}{d \ln \sigma} \right]
- \left[ S_3 (\sigma) + U_3 (\sigma) + V_3 (\sigma) + \frac {d S_3}{d \ln \sigma} + \frac {d U_3}{d \ln \sigma}  \right] ,
\ee
where $U_3$ and $V_3$ are given by:
\ba
U_3 (\sigma) &=& \frac {3}{\sigma^2} \left[ \frac {d}{d(\sigma_1^2)} \langle \delta (\sigma_1^2) \delta^2 (\sigma^2) \rangle \right]_{\sigma_1^2 = \sigma^2} \\
V_3 (\sigma) &=& \frac {9}{2} \left[ \frac{d^2}{d(\sigma_1^2)^2} \langle \delta (\sigma_1^2) \delta^2 (\sigma^2)  \rangle \right]_{\sigma_1^2=\sigma^2} 
+ 12 \left[ \frac {d}{d(\sigma_1^2)} \frac {d}{d(\sigma_2^2)}\langle \delta (\sigma_1^2) \delta (\sigma_2^2) \delta (\sigma^2) \rangle  \right]_{\sigma_1^2=\sigma_2^2=\sigma^2}\;.
\ea
The MR function does not need any ad-hoc rescaling.

\paragraph*{\bf Lam-Sheth (LS)}

An extension of the ST model to primordial non-Gaussianity
has been recently proposed by \cite {2009arXiv0905.1702L}.
In this case the mass function is written as:
\be
f_{\mathrm {LS}} \left( \frac{\delta_c}{\sigma} \right) = f_{\mathrm {ST}} \left( \frac{\delta_c}{\sigma} \right) \left\{ 1 + \frac{\sigma S_3}{6} H_3 \left[ \frac {b (\sigma)}{\sigma}  \right]  \right\},
\ee
where $H_3 \left(x \right) = x (x^2 - 3)$, and the mass-dependent collapse
barrier is 
$b(\sigma) = \sqrt{a} \delta_c \left[ 1 + \beta \left( \sigma / \sqrt {a} \delta_c \right)^{2 \gamma} \right]$, with $\beta = 0.4$, $\gamma = 0.6$, $a = 0.7$.
Note that, in the case of a constant barrier, this mass function reduces to the
LV one, if in the latter 
we neglect the term proportional to the derivative of $S_3$, which
is generally small.

\subsection{Comparison with N-body simulations}
In Fig.~\ref {fig:mf} we compare the theoretical mass functions presented
above with the N-body data by PPH08.
The halos in the simulations were identified using a friends-of-friends algorithm with linking
length $b = 0.2 \lambda$, where $\lambda$ is the mean interparticle distance.
Consistently with PPH08,
for our calculations
we assume the WMAP5 $\Lambda$CDM model, with parameters $h=0.701, \sigma_8=0.817, n_s = 0.96, \Omega_m = 0.279, \Omega_b = 0.0462, \Omega_{\Lambda} = 0.721$.
We find that
all the theoretical mass functions match the numerical output  within
$\sim 10-20\%$ for $\fnl<500$. Most of the discrepancy originates from the
fact that the ST mass function underestimates the halo counts from the simulations
(left panel in Fig.~\ref {fig:mf}). Indeed, the models are rather accurate in
predicting the ratio between the counts in a non-Gaussian model vs. a Gaussian
one (see also \cite {Desjacques:2008vf, Pillepich:2008ka, Grossi:2009an}).
We also consider
a fitting formula for the mass function that was computed by PPH08 
from the very same data plotted in Fig.~\ref {fig:mf}.
Here we rewrite it as
\be
f_{\mathrm {PPH}} \left( \frac {\delta_c} {\sigma} \right) = \left[ D + B \left( \frac{\delta_c}{1.686 \, \sigma} \right)^A  \right] \exp \left( - \frac{C \delta_c^2}{1.686^2 \, \sigma^2} \right),
\ee
 where we have explicitly introduced a dependence on the threshold collapse
 density $\delta_c$, 
and $A, B, C, D$ are fitting parameters which depend on $\fnl$. We use the
values from Table~5 in PPH08.
 
In the next section, we will use the mass function to compute
the halo bias parameters in the non-Gaussian scenario. Given that all the
models for $n$ are of the same quality, as a reference, we will only show the results obtained
with the LV mass function and the PPH fit.

\section {Halo bias} \label {sec:bias}

\subsection {Peak-background split} \label{sec:pbsplit}
We decompose the Gaussian auxiliary potential $\varphi$ into the
(statistically independent) contributions of long- and short-wavelength modes:
\be
\varphi (\mathbf{q}) = \varphi_l(\mathbf{q}) + \varphi_s (\mathbf{q})\;.
\ee
Eq.~(\ref{eq:basic}) then gives
\begin{eqnarray} \label{eq:phis}
\Phi_l&=& \varphi_l + \fnl  \varphi_l^2-\langle \varphi^2 \rangle \nonumber\\
\Phi_{m}&=& 2 \fnl \varphi_l \, \varphi_s \\
\Phi_s&=& \varphi_s + \fnl \varphi_s^2 \;, \nonumber
\end{eqnarray}
where the dependence on the spatial position is understood.
The mixed term $\Phi_{m}$  contributes to the short-wavelength part but derives
from the coupling of $\varphi_l$ and $\varphi_s$.
It vanishes for Gaussian initial conditions.
When passing from real to Fourier space, the products of two fields become convolutions. Strictly speaking, the terms $\varphi_l \varphi_s$ and $\varphi_s^2$ would also contribute to the long modes $\Phi_l$, due to the mixing of modes caused by the convolution operation. We have checked however that these additional contributions are completely subdominant.

Using the Poisson equation, $\nabla^2 \Phi=A \, \delta$ with $A = 3 \Omega_m H_0^2 /
(2 c^2)$,  for the density fluctuations we can then write
\begin{eqnarray} \label{eq:deltas}
\delta_l&=& \delta_{Gl} (1+2\fnl \varphi_l) +2 A^{-1} \fnl \nabla \varphi_l \cdot \nabla
\varphi_l \nonumber\\
\delta_{m}&=& 2\fnl (\delta_{Gs} \,\varphi_l+\delta_{Gl}\, \varphi_s)+4 A^{-1} \fnl \nabla \varphi_l \cdot \nabla
\varphi_s \nonumber \\
\delta_s&=& \delta_{Gs} (1+2\fnl \varphi_s)+2 A^{-1} \fnl \nabla \varphi_s \cdot \nabla
\varphi_s\;, 
\end{eqnarray}
where $\nabla^2 \varphi=A \, \delta_{G}$. Notice that:
\be \label{eq:venti}
\delta_m = 2\fnl\left[\frac{\delta_s-2A^{-1}\fnl \nabla \varphi_s \cdot \nabla
\varphi_s}{1+2\fnl \varphi_s}\,\varphi_l\,+
\frac{\delta_l-2A^{-1}\fnl \nabla \varphi_l \cdot \nabla
\varphi_l}{1+2\fnl \varphi_l}\,\varphi_s \right]
+4 A^{-1} \fnl \nabla \varphi_l \cdot \nabla\varphi_s
\;.
\ee

In the spirit of the peak-background split \cite {Bardeen:1985tr, Cole:1989vx, Mo:1995cs, Catelan:1997qw},
the short-wavelength modes of the density field collapse to form virialized 
condensations (dark-matter halos) while the long-wavelength ones modulate 
the halo counts and are responsible for large-scale motions.
In the non-Gaussian case, however, the halo collapse will also be influenced
by the long-wavelength modes of $\varphi$ and $\nabla \varphi$ 
which contribute to $\delta_m$.
In a Press-Schechter approach,
modulations in $\delta_l$ will modify the threshold for halo collapse
as in the Gaussian case.
However, \emph {in the presence of non-Gaussian fluctuations, the large-scale
modes of the pseudo-potential will also alter the statistical properties
of the small-scale modes in the density field}. This provides an additional
source of biasing with respect to the Gaussian case.
Suppose we want to apply the Press-Schechter algorithm to $\delta$.
The probability that the small-scale fluctuation $\delta_s+\delta_m$
is above the collapse threshold $\delta_c-\delta_l$ (probability which is obtained
by averaging over $\delta_s$)
would then explicitly depend on $\delta_l$, $\varphi_l$ 
and $\nabla \varphi_l$.
This implies that the resulting large-scale halo overdensity cannot
be proportional to $\delta_l$ as in the Gaussian case (to first order).
Rather, in the general case, $\delta_h\simeq b_1 \delta_l+f_1 \varphi_l
+g_1 \nabla \varphi_l\cdot \nabla \varphi_l$ plus higher-order terms.
The bias coefficients 
are given by the Taylor expansion of the conditional mass function
$n(M|\delta_l,\varphi_l, \nabla \varphi_l\cdot \nabla \varphi_l)$. 
Unfortunately,
the models for the mass function listed in the previous section have been
obtained by averaging over the entire Lagrangian volume and have no memory
of the cross-talk between large and small scales.
We attempted the calculation of 
$n(M|\delta_l,\varphi_l, \nabla \varphi_l\cdot \nabla \varphi_l)$ by adopting a Press-Schechter
approach and starting from the Gaussian fields $\varphi, \nabla \varphi$ and 
$\nabla^2 \varphi$ but we could not obtain a closed form
due to the complexity of the expressions. 

An approximated model can be obtained assuming that
halos form from the highest peaks of $\delta_s$.\footnote{We only assume that halo formation happens around some of the density peaks. This is different from the approach by \cite{Matarrese:2008nc,Desjacques:2009jb}, in which all peaks form halos, and a one-to-one correspondence between them is assumed.}
We want to implement this requirement in Eq.~(\ref{eq:deltas}).
 Let us consider what the labels "short" and "long" mean in practical terms.
The short part of the fields will only include a narrow shell of modes
centered around the wavelength corresponding to Lagrangian size of the
halos. On the other hand, the long part will be formed with all the
Fourier modes with larger wavelengths.
In this case, $\varphi_s$ will be closely tracing $\delta_{Gs}\simeq
\delta$. This implies that the
high  density peaks will nearly coincide with the maxima of
$\varphi_s$, where $\nabla \varphi_s=0$. In this case,
\be
\delta_s+\delta_m=\delta_s
\left(1+\frac{2\fnl \varphi_l}{1+2\fnl \varphi_s}\right)+
\frac{\delta_l-2A^{-1}\fnl \nabla \varphi_l \cdot \nabla
\varphi_l}{1+2\fnl \varphi_l}\,2\fnl \varphi_s
\;. \nonumber
\ee
The Lagrangian size of galaxy- and cluster-sized halos ranges between
1 and 10 Mpc. 
This implies that $\langle \delta_s^2 \varphi_l^2 \rangle^{1/2} \gg
\langle \delta_l^2 \varphi_s^2 \rangle^{1/2}$
and  $\langle \varphi_l^2 \rangle \simeq \langle \varphi_s^2\rangle$, 
since perturbations in the pseudo-potential are nearly scale invariant.
Moreover, $\fnl \langle \varphi_l^2 \rangle^{1/2}\ll 1$ 
for the values of $\fnl$ of physical interest.
We thus obtain
\be \label{eq:dng}
\delta_s+\delta_m\simeq \delta_s
\left(1+2\fnl \varphi_l\right)\;,
\ee
i.e. the amplitude of small-scale density fluctuations is enhanced
in regions where $\varphi_l$ is large.
Therefore, the conditional mass function $n(M|\varphi_l)$  can be computed
from the unconditional one $n(M)$ by simply multiplying the r.m.s.
of the density fluctuations by the factor $1+2\fnl
\varphi_l$.\footnote{\citet{Slosar:2008hx} 
and \citet{Afshordi:2008ru} derived a similar expression but notice
that ours is written in terms of the non-Gaussian density field.}
At the same time, we can use the peak-background split to derive 
$n(M|\delta_l,\varphi_l)$ by simply replacing $\delta_c$ with 
$\delta_c-\delta_l$.

In each point in Lagrangian space, we can thus define a Lagrangian halo density field as
\be 
\delta^L_h(\mathbf{q}) = \frac {n [M, \delta_l(\mathbf{q}), \varphi_l(\mathbf{q})]}{\bar n} - 1,
\ee
where the average can simply be taken as $\bar n = n (M, 0, 0)$. Here 
it is possible to
replace $n$ with $f$ 
since the proportionality factors cancel out, so that we can write more explicitly
\be \label{eq:moreexplic}
\delta^L_h(\mathbf{q}) = \frac {f \left( \frac{\delta_c - \delta_l(\mathbf{q})}{\left[ 1 + 2 \fnl \varphi_l(\mathbf{q}) \right] \sigma}   \right) }{f \left( \frac{\delta_c}{\sigma} \right)} - 1.
\ee
We can then expand the perturbations in a Taylor series in terms of \emph{both} variables
$\delta_l$ and $\varphi_l$, obtaining
\be \label{eq:deltah0}
\delta_h^L (\mathbf{q}) =  \sum_{j=0}^{\infty} \sum_{m=0}^{\infty} \frac
{b_{jm}^L}{j!m!} \, \delta_l^j (\mathbf{q}) \, \varphi_l^m (\mathbf{q}) \, .
\ee
Up to third order in the perturbations, this gives:
\ba \label{eq:bfexp0}
\delta^L_{h} (\mathbf q) &=& 
b^L_0 + b^L_{10} \, \delta + b^L_{01}\, \varphi + \nonumber \\
&+& \frac{1}{2!} \left( b^L_{20} \, \delta^2 + 2\,b^L_{11}\, \delta  \varphi + b^L_{02} \, \varphi^2 \right)+ \nonumber \\
&+& \frac {1} {3!} \left( b^L_{30}\, \delta^3 + 3 \, b^L_{21} \, \delta^2 \varphi + 3 \, b^L_{12}\, \delta \varphi^2 + b^L_{03}\, \varphi^{3} \right),
\ea
where all the density perturbations on the r.h.s. are Lagrangian and
non-Gaussian.

Eq.~(\ref{eq:moreexplic}) implies that not all the coefficients
$b_{jm}^L$ are independent. In particular, all the $b_{jm}^L$
with $m \ne 0$ can be written in terms of the $b_{j0}^L$. 
Up to third order we have: 
\ba \label{eq:bsaintindep}
b_{01}^L &=& 2 \, \fnl \, \delta_c \, b_{10}^L \nonumber \\
b_{11}^L &=& 2 \, \fnl \, ( - b_{10}^L + \delta_c \, b_{20}^L) \nonumber \\
b_{02}^L &=& 4 \, \fnl^2 \, (- 2 \delta_c \, b_{10}^L + \delta_c^2 \, b_{20}^L )  \nonumber \\
b_{21}^L &=& 2 \, \fnl \, (- 2 b_{20}^L + \delta_c \, b_{30}^L ) \nonumber \\
b_{12}^L &=& 4 \, \fnl^2 \, ( 2 b_{10}^L - 4 \delta_c \, b_{20}^L + \delta_c^2 \, b_{30}^L ) \nonumber \\
b_{03}^L &=& 8 \, \fnl^3 \, ( 6 \delta_c \, b_{10}^L - 6 \delta_c^2 \, b_{20}^L   + \delta_c^3 \, b_{30}^L ).
\ea
It is important to remember that
the functional form of the halo mass function accounts for the effect of
non-Gaussianity on the short wavelength modes $\delta_s, \delta_m$. The bias
coefficients $b_{j0}^L$ will then depend implicitly on $\fnl$ through the shape of the mass function.

\subsection {Extension to higher-order non-Gaussianity} \label{sec:gnl}

If the model for non-Gaussianity is extended to higher order, then
Eq.~(\ref{eq:basic}) is replaced by Eq.~(\ref{eq:basichi}). If we consider
cubic corrections,
we have the following additional contributions  to Eqs.~(\ref{eq:phis}):
\begin{eqnarray}
\Delta \Phi_l&=& \gnl \, \varphi_l^3 \nonumber\\
\Delta\Phi_{m}&=& 3 \gnl  \left( \varphi_l^2 \, \varphi_s + \varphi_l  \, \varphi_s^2 \right) \\
\Delta \Phi_s&=& \gnl \, \varphi_s^3 , \nonumber
\end{eqnarray}
which correspond to the following additions in Eqs.~(\ref{eq:deltas}):
\begin{eqnarray}
\Delta \delta_l&=& 3 \gnl \, \varphi_l^2 \, \delta_{Gl} + 6 A^{-1} \gnl \, \varphi_l \, \nabla \varphi_l \cdot \nabla
\varphi_l \nonumber\\
\Delta \delta_{m}&=& 3 \gnl \, \delta_{Gl} \,  \varphi_s (2  \varphi_l +  \varphi_s) + 6 \gnl A^{-1} \, \varphi_s \, \nabla \varphi_l \cdot \nabla \varphi_l  \\
\Delta \delta_s&=& 0 \nonumber,
\label{eq:deltashi}
\end{eqnarray}
where we have already imposed the peak condition.
In analogy with the previous section we thus identify the leading term as:
\be \label{eq:gnl}
\delta_s+\delta_m \simeq \delta_s \left( 1 + 2 \fnl \varphi_l + 3 \gnl \varphi_l^2  \right).
\ee
It follows that the r.m.s. of the small-scale density fluctuations will now be altered by
a factor 
$\left( 1 + 2 \fnl \varphi_l + 3 \gnl \varphi_l^2  \right)$ with respect to the Gaussian case.
Therefore, considering $\gnl \ne 0$ introduces additional terms in the
bias coefficients 
in Eq.~(\ref{eq:bsaintindep}), given by
\ba \label{eq:dbgnl}
\Delta b_{02}^L &=& 6 \, \gnl \, \delta_c \, b_{10}^L   \nonumber \\
\Delta b_{12}^L &=& 6 \, \gnl \, ( - b_{10}^L + \delta_c \, b_{20}^L),
\ea
while all the other bias coefficients remain unchanged (apart from the
modifications due to the implicit dependence of the mass function on $\gnl$, which we do not calculate here).

Eq.~(\ref{eq:gnl}) can be finally generalized to an arbitrary order $N$ as
\be
\delta_s+\delta_m \simeq \delta_s \sum_{j=2}^N j Q_{\mathrm{NL}j} \varphi_l^{j-1}.
\ee
This equation shows that the leading contribution of each successive
order will depend on a higher power of the potential, and its
effects will therefore be smaller in amplitude. Note that 
$h_{\mathrm{NL}} \equiv Q_{\mathrm{NL}4}$ is the highest-order term 
that can explictly modify 
the $b_{ij}$ parameters (up to third order in the bias expansion), 
although all the $ Q_{\mathrm{NL}j}$ will
introduce implicit dependences by modifying the halo mass function.

\subsection {Bias from a mass function}
We want now to explicitly calculate the halo bias corresponding to a given mass function.
As discussed above, in the non-Gaussian case the mass function will also be
explicitly dependent on the potential $\varphi$, and the halo overdensities
can now be derived from  Eq.~(\ref{eq:deltah0}), as a bivariate series expansion in
terms of $\delta_l$ and $\varphi_l$. Since the effect of the short-wavelength modes
is taken into account by 
the functional form of the mass function, we will henceforth drop the $l$
indices and use the symbols $\delta$ and $\varphi$ to denote the
long-wavelength parts of the perturbations.

\begin {figure}
\begin{center}
\includegraphics[width=0.45\columnwidth,angle=0]{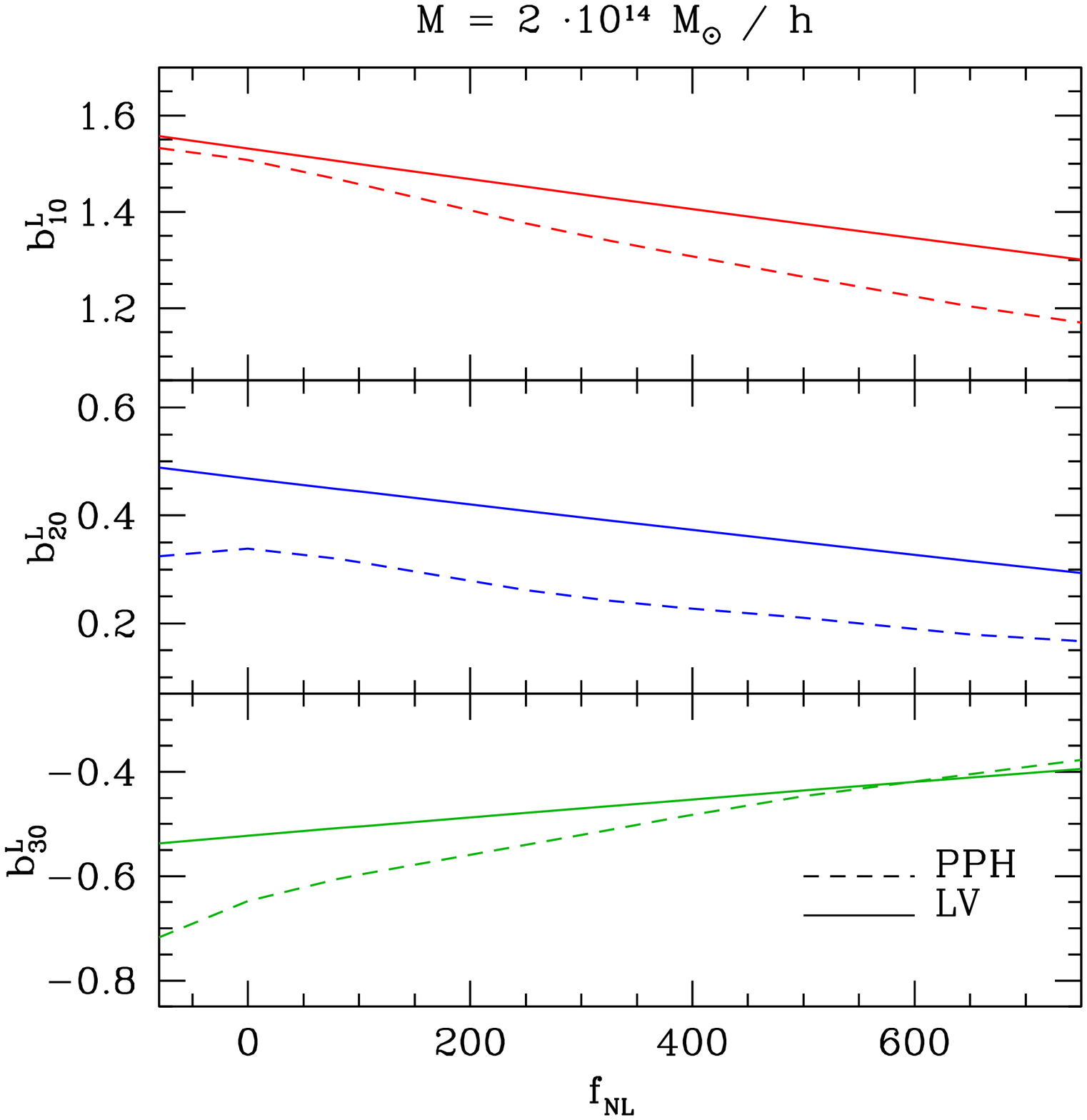}
\includegraphics[width=0.45\columnwidth,angle=0]{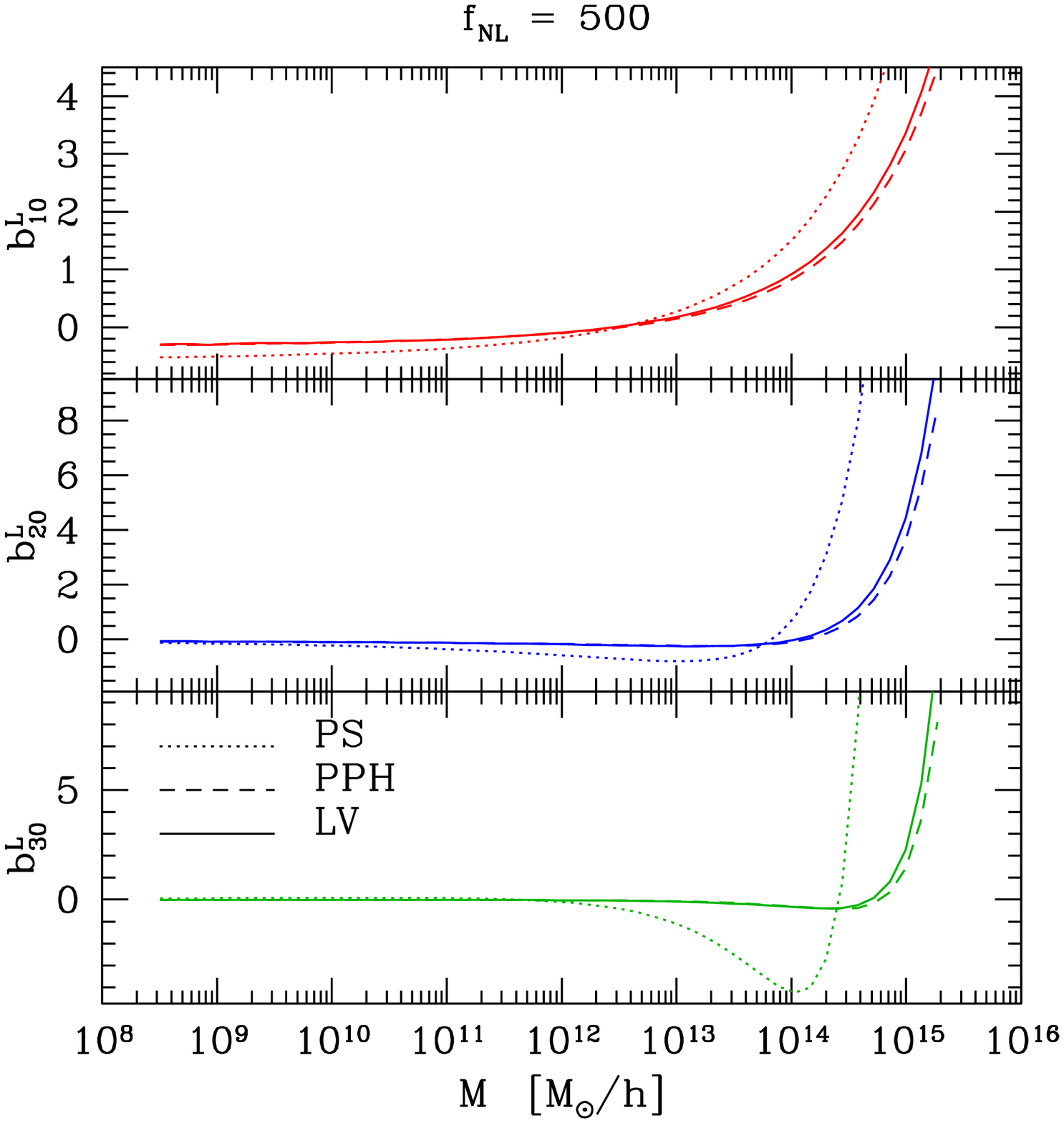}
\end{center}
\caption{Lagrangian bias factors at $z=0$ as a function of $\fnl$ for a halo
  mass of $2 \cdot 10^{14} M_{\odot}/h$ (left), and
 as a function of the halo mass, for $\fnl = 500 $ (right). The results
 obtained from the PPH and LV mass functions are shown in both cases. For
 reference, in the right panel we also show the results obtained from the PS
 mass function, which does not depend on $\fnl$.}
\label{fig:b1b2b3}
\end{figure}

\paragraph*{\bf PS model} 
For reference, 
we derive the bias coefficients corresponding to the simple PS mass function
(see also \cite{Mo:1995cs})
\ba \label {eq:b1}
b^L_{10} &=& -\frac{1}{\delta_c} + \frac {\delta_c}{\sigma^2},  \nonumber \\
b^L_{20} &=& \frac {\delta_c^2}{\sigma^4} - \frac{3}{\sigma^2} \nonumber \\
b^L_{30} &=& \frac{\delta_c^3}{\sigma^6} - \frac{6 \delta_c}{\sigma^4} + \frac{3}{\delta_c \sigma^2}
\ea
and show their mass dependence in the right panel of Fig.~\ref{fig:b1b2b3}.
The remaining coefficients can be obtained using Eq.~(\ref{eq:bsaintindep}): 
\ba
b^L_{01} &=& \left( \frac {2 \delta_c^2 }{\sigma^2} - 2 \right) \fnl  \nonumber \\
b^L_{11} &=&   \left( \frac{2}{\delta_c} +  \frac {2 \delta_c^3}{\sigma^4} -  \frac {8 \delta_c}{\sigma^2} \right) \fnl  \nonumber \\ 
b^L_{02} &=&  2 \left( \frac{ 2 \delta_c^4}{\sigma^4} - 10 \frac {\delta_c^2} {\sigma^2} + 
   4  \right) \fnl^2  \nonumber \\ 
b^L_{21} &=& 2 \left(\frac {\delta_c^4}{ \sigma^6} - \frac {8 \delta_c^2}  {\sigma^4} + \frac {9} {\sigma^2} \right) \fnl  \nonumber \\
b^L_{12} &=& 2 \left( \frac {2 \delta_c^5}{ \sigma^6} - \frac {20 \delta_c^3 }{ \sigma^4} + \frac {34 \delta_c }{ \sigma^2} - \frac {4}{ \delta_c}  \right) \fnl^2    \nonumber \\
b^L_{03} &=& 6 \left( \frac {4 \delta_c^6}{3  \sigma^6} - \frac{16 \delta_c^4 }{  \sigma^4} + \frac{36 \delta_c^2 }{  \sigma^2} - 8  \right)  \fnl^3.   
\ea
Notice that the ``usual'' bias coefficients ($b^L_{10}, b^L_{20}, b^L_{30}$)
are in this case independent 
from $\fnl$. This does not hold in general, since an implicit dependence on
$\fnl$ will be introduced by any 
non-Gaussian mass function. 

\paragraph*{\bf General case} 
To derive the bias coefficients up to the third order,
we can repeat the same procedure for any other mass function.
We have calculated these coefficients for all the mass functions listed in 
Section \ref{sec:massfunc} finding an overall agreement in the trends with
mass and with the non-linearity parameter $\fnl$.
The analytic form of the bias parameters is much more complex
than for the PS mass function and we will not write it explicitly.
As an example, 
in Fig.~\ref{fig:b1b2b3} we show how the bias coefficients $b_{j0}^L$ depend
on $\fnl$ and halo mass for the LV mass function and the PPH fit.

\subsection {Lagrangian and Eulerian bias}

A model for the formation of the LSS provides a relationship 
between the density perturbations in Lagrangian and Eulerian space.
This relation is generally non-local \cite{Catelan:1997qw} but a local approximation (where $\mb x \equiv \mb q$) suffices to
approximately describe the evolution of large-scale perturbations. In this case
one writes \cite{Mo:1995cs,Mo:1996zb}
\be \label {eq:series}
\delta^L = \sum_{j=1}^{\infty} a_j \left( \delta^E \right)^j,
\ee
where the $a_j$'s parameterize the evolution of mass-density fluctuations. 
For the simple case of spherical collapse, we have \cite{Bernardeau:1994zd}
\be
a_1 = 1 \, ; \:\: a_2 = -17 / 21 \, ; \:\: a_3 = 341 / 567.
\ee

Starting from the Lagrangian halo density perturbations $\delta_{h}^L$  given in Eq.~(\ref{eq:bfexp0})
we want to use Eqs.~(\ref{eq:eulag4}) and (\ref{eq:series}) to write the
Eulerian halo overdensity in terms of Eulerian density perturbations. This gives:
\ba \label{eq:bfexpEuler}
\delta_{h} (\mathbf x) &=& b_0 + b_{10} \, \delta + b_{01}\, \varphi+ \nonumber \\
&+& \frac{1}{2!} \left(b_{20} \, \delta^2 + 2 \, b_{11} \, \delta \varphi + b_{02} \, \varphi^2  \right) + \nonumber \\
&+& \frac{1}{3!} \left( b_{30} \, \delta^3 + 3 \, b_{21} \, \delta^2 \varphi + 3 \, b_{12} \, \delta \varphi^2 + b_{03} \, \varphi^{3} \right),
\ea
where all the density perturbations on the r.h.s. are Eulerian and
non-Gaussian
(from now on we drop the superscript $E$ and all densities will be
Eulerian unless explicitely stated otherwise) and
the bias coefficients are given by the following expressions:
\ba \label{eq:eubi}
b_{10} &=&  1 + a_1 \, b^L_{10} \nonumber   \\
b_{20} &=&  2 (a_1 + a_2) \, b^L_{10} + a_1^2 \, b^L_{20} \nonumber \\ 
b_{30} &=&  6 (a_2 + a_3) \, b^L_{10} +  3 \left(  a_1^2 + 2 a_1 a_2 \right) \, b^L_{20} + a_1^3 \, b^L_{30}
\\
b_{01} &=&  b^L_{01}  \nonumber \\
b_{11} &=&  b^L_{01} + a_1 \,  b^L_{11} / 2  \nonumber \\ 
b_{02} &=&  b^L_{02} \nonumber \\
b_{21} &=&  (a_1 + a_2) \, b^L_{11}  + a_1^2 \, b^L_{21} / 3 \nonumber \\ 
b_{12} &=&  b^L_{02} + a_1 \, b^L_{12} / 3 \nonumber \\ 
b_{03} &=&  b^L_{03}.  \nonumber
\ea
Note that Eq.~(\ref{eq:bfexpEuler}) differs from Eq.~(\ref{ldb}) due to
the presence of extra terms which are proportional to different powers
of the Gaussian auxiliary potential $\varphi$.
Since $\varphi$ is nearly scale invariant while $\delta$ has a larger
variance on smaller scales,
the additional terms will only affect the statistics of the halo
distribution on the largest scales.
Also, $\varphi$ does not evolve with time while $\delta$ does (according to
$\delta(z) = D(z) \delta(z=0)$ 
at linear order) thus implying
that, for a given set of bias coefficients, the new terms will become less and
less important over time. 
A third peculiarity of  Eq.~(\ref{eq:bfexpEuler}) is that the halo
overdensity can differ from zero also in regions with mean mass density.

\subsection {Perturbative expansion}
In order to account for the non-linear evolution of mass-density fluctuations 
in Eq.~(\ref{eq:bfexpEuler})
we use standard Eulerian perturbation theory (see \cite{Bernardeau:1994zd} for a
 review). We therefore
expand the density fields to third order as $\delta = \delta_{1} +
\delta_{2} + \delta_{3} + {\cal O}(\delta_4) $ where $\delta_n$ is ${\cal
  O}(\delta_1^n)$. Each term can be expressed as
\be
\tilde
\delta_n(\mb k)=\int \frac{d^3 \mb q_1}{(2 \pi)^3} \dots \frac{d^3 \mb q_n}{(2 \pi)^3} 
\,\delta_D\left(\mb k-\sum_{i=1}^n \mb
q_i\right)\,
J_n(\mb q_1,\dots,\mb q_n)\, \tilde \delta_1(\mb q_1) \dots \tilde
\delta_1(\mb q_n)\;,
\label{eq:perturbn}
\ee
where $\delta_D$ is the Dirac delta distribution,
the tilde denotes Fourier transformation, and the $J_n$ are specific
kernel functions. 
On the other hand, since the Gaussian potential $\varphi$ is the primordial one, 
there is no need to expand it, and it fully coincides with its first-order part $\varphi \equiv \varphi_{1} $. 

We can now explicitly rewrite Eq.~(\ref{eq:bfexpEuler}) up to the third
perturbative order as
\ba \label{eq:deltaE}
\delta_{h} (\mathbf x) &=& b_0 + q_{10} \delta_{1}  + q_{11} \varphi_{1}  + \nonumber  \\
&+& 
q_{20} \delta_{2} + 
q_{21} \delta_{1}^2   + 
q_{22} \delta_{1} \varphi_{1} + 
q_{23} \varphi_{1}^2  + \nonumber  \\
&+&
q_{30} \delta_{3} +
q_{31} \delta_{1} \delta_{2} +
q_{32} \delta_{2} \varphi_{1} + 
q_{33} \delta_{1}^2 \varphi_{1} + 
q_{34} \delta_{1} \varphi_{1}^2 + 
q_{35} \delta_{1}^3  +
q_{36} \varphi_{1}^3 ,
\ea
where, to simplify the notation and facilitate bookkeeping of the terms which
will appear in the perturbative expression for the power spectrum, we have replaced
the biases $b_{ij}$ with new coefficients $q_{ij}$.
The explicit form of these  is given in Table~\ref{tab:coef}. 

Eq.~(\ref{eq:deltaE}) fully describes the Eulerian halo bias 
at the third perturbative order in the non-Gaussian case.
The leading order term,  $\delta_h\simeq [1+a_1b_{10}^L (\fnl) ] \, \delta+
2 \fnl b_{10}^L (\fnl) \, \varphi$,
was already recognized by Dal07, \cite{Afshordi:2008ru} and
\cite{Slosar:2008hx}.

\begin{table}
\begin{tabular}{|c| c|  c|  c| c|  c|  c|}
\hline
$q_{10} = b_{10}$  &  $q_{11} = b_{01}$  &   & &   &    &   \\
$q_{20} = b_{10}$  &  $q_{21} = b_{20} / 2 $  & $q_{22} = b_{11} / 2 $   & $q_{23} = b_{02} / 2 $  &   &    &   \\
$q_{30} = b_{10}$  &  $q_{31} = b_{20}$  &  $q_{32} = b_{11} / 2$ &  $q_{33} = b_{21} / 6$ &  $q_{34} = b_{12} / 6$  & $q_{35} = b_{30} / 6$   & $q_{36} = b_{03} / 6$  \\
\hline
\end{tabular}
\caption{Mapping of the bias coefficients in the full perturbative expansion.}
\label{tab:coef}
\end{table}

\section {Clustering statistics and non-Gaussianity} \label{sec:statistics}

Statistical analysis of random fields, such as the  Bardeen
potential $\Phi (\mb x)$, can be performed by studying the 
irreducible $N$-point correlation functions
$\langle \Phi (\mb x_1) \Phi (\mb x_2) ... \Phi (\mb x_N) \rangle$,
or alternatively the $N$-spectra
\be
(2 \pi)^3 {\cal S}_N (\mathbf k_1,\mathbf k_2, ..., \mathbf k_N) \,  \delta_D (\mathbf k_1 + \mathbf k_2 + ... + \mathbf k_N) = \langle \tilde \Phi (\mathbf k_1) \tilde \Phi (\mathbf k_2) ... \tilde \Phi (\mathbf k_N)  \rangle\;.
\ee
For Gaussian fields all odd-order spectra vanish. On the other hand, thanks to Wick's
theorem, the reducible even-order correlators can be decomposed as products of the power
spectrum, 
\be
(2 \pi)^3 P_{\Phi}(k) \, \delta_D (\mathbf k_1 + \mathbf k_2) = \langle \tilde  \Phi (\mathbf k_1) \tilde  \Phi (\mathbf k_2) \rangle,
\ee
which, in this case, encodes all the information. 
If some non-Gaussianity is instead introduced, then higher-order statistics
become important, such as the bispectrum 
\be
(2 \pi)^3 B_{\Phi}(\mathbf k_1, \mathbf k_2, \mathbf k_3) \, \delta_D (\mathbf k_1 + \mathbf k_2 + \mathbf k_3) = \langle \tilde  \Phi (\mathbf k_1) \tilde  \Phi (\mathbf k_2) \tilde  \Phi (\mathbf k_3) \rangle,
\ee
and the irreducible trispectrum 
\be
(2 \pi)^3 T_{\Phi}(\mathbf k_1, \mathbf k_2,\mathbf k_3,\mathbf k_4) \, \delta_D (\mathbf k_1 + \mathbf k_2 + \mathbf k_3 + \mathbf k_4) = \langle \tilde  \Phi (\mathbf k_1) \tilde \Phi (\mathbf k_2) \tilde \Phi (\mathbf k_3) \tilde \Phi (\mathbf k_4) \rangle.
\ee
Using our simplest model of non-Gaussianity given in Eq.~(\ref{eq:basic})
to define the non-Gaussian potential $\Phi$ in terms of the Gaussian one 
$\varphi$, we obtain:
\be \label{eq:simplifme}
P_{\Phi}(k) = P_{\varphi}(k) + 2 \fnl^2 \int \frac{d^3 \mb{q}}{(2 \pi)^3}
P_{\varphi}(q) P_{\varphi}(|\mb{k} - \mb{q}|)\simeq P_{\varphi}(k)
\ee
\be
B_\Phi(\mathbf k_1,\mathbf k_2,\mathbf k_3) \simeq 2 \fnl \left[P_\varphi(k_1) P_\varphi(k_2) + \mathrm {(2\ cyclic)}\right]
\label{eq:bispPhi}
\ee
\be
T_{\Phi}(\mathbf k_1, \mathbf k_2, \mathbf k_3, \mathbf k_4) \simeq 4 \fnl^2 \left\{P_\varphi(k_1)
P_\varphi(k_2) \left[P_\varphi(|\mb k_1+\mb k_3| )+ P_\varphi(|\mb
  k_1+ \mb k_4|)\right]+\mathrm{ (5\ cyclic)}\right\}
\label{eq:trispPhi}
\ee
where we dropped a sub-leading term proportional to $\fnl^N$ for each of the
$N$-spectra (this is why we used the symbol of approximate
equality).\footnote{In multi-field inflationary models the
  non-Gaussian contribution to the trispectrum may scale independently from
  the bispectrum. For this reason, the factor $\fnl^2$ in
  Eq.~(\ref{eq:trispPhi}) is sometimes re-labeled $\tau_{\mathrm{NL}}$, and
  treated as an independent parameter. Observational constraints on the
  trispectrum of the Bardeen's potential should then discriminate between such models and the simplest inflationary scenarios.}
We have checked that the discarded terms
are indeed negligibly small.
For instance, the sub-leading contribution to $P_\Phi$ contributes less
than $1\%$ of the total for $|\fnl| \sim 1000$ in the $k$-range of interest.

Considering also the third-order term in Eq.~(\ref{eq:basichi})
introduces additional contributions 
to the power spectrum of the Bardeen's potential. The leading-order term
can be written as:
\be 
\Delta P_{\Phi}(k) =  6 \gnl P_{\varphi}(k) \int \frac{d^3 \mb{q}}{(2 \pi)^3}
P_{\varphi}(q)\;. 
\ee
For  $P_{\varphi}(k)\propto k^{n_s-4}$, the integral above presents
an ultraviolet $(k \to \infty)$ divergence if $n_s\geq 1$ and
an infrared $(k \to 0)$ divergence if $n_s\leq 1$. 
In general, this is not a problem as 
the physical process creating the fluctuations  
will automatically introduce cutoffs in $P_\varphi$
at small and large wavelengths.
For example, cosmic inflation will generate perturbations 
with characteristic sizes comprised between the reheating scale
and the present-day horizon
\cite{Matarrese:2000iz}.
However, if $\Delta P_\Phi$ is non-negligible with respect to the
leading-order contribution $P_\varphi$ (i.e. if $6|\gnl| \langle \varphi^2
\rangle$ is not much less than unity), the results of the 
perturbative expansion are of limited use unless artificial cutoffs are 
introduced and the parameters of the theory are renormalized.
The condition above reduces to $|\gnl| \ll 10^7$ if the currently favored
values for the amplitude and the spectral index of primordial perturbations 
are plugged in. Present-day observational limits on $\gnl$ 
\cite{Desjacques:2009jb,Vielva:2009jz}
therefore suggests
that $\Delta P_\Phi$ should contribute at the percent level or less to
the power spectrum of the potential.
Note that in numerical simulations 
\cite{Desjacques:2009jb}, non-physical 
infrared and ultraviolet cutoffs are introduced
by the use of a finite volume with periodic boundary conditions.
Considering a non-vanishing $\gnl$ also adds
another leading-order correction to the trispectrum of the Bardeen potential:
\be \label{eq:deltaT}
\Delta T_{\Phi}(\mathbf k_1, \mathbf k_2, \mathbf k_3, \mathbf k_4)\simeq 6 \gnl P_\varphi(k_1) P_\varphi(k_2)
P_\varphi(k_3) + \mathrm{(3\ cyclic)}\;, 
\ee
while it does not modify the bispectrum of $\Phi$ at leading order.

Linear perturbations in the density at redshift $z$ are related to those in 
the primordial potential (formally at $z\to \infty$)
by the Poisson equation
\be \label{eq:poissalpha}
\tilde \delta_1(k) = \alpha(k) \tilde  \Phi(k) 
\ee
with
\be
\alpha(k) = \frac {2 c^2 k^2 T(k) D(z)} {3 \Omega_m H_0^2} \frac{g(0)}{g(\infty)},
\ee
where the matter growth factor $D(z)$ and the transfer function $T(k)$ have been
introduced to account for the linear evolution of $\delta_1$. The function
$g(z)\equiv (1+z)D(z)$ is the linear growth factor for the potential,
and $g(\infty)/g(0)\simeq 1.3$ in the currently favored cosmology.
Therefore, we can relate the power spectrum of linear density fluctuations 
to the power spectrum of the primordial potential by writing
\be
P_0 (k) \equiv P_{\delta_1}(k) = \alpha^2(k) \, P_{\Phi} (k)  \simeq  \alpha^2(k) \,
P_{\varphi} (k),
\label{eq:convspectra}
\ee
where the last approximation follows from Eq.~(\ref{eq:simplifme}). Similar
equations can be written for the three- and four-point correlators of the linear density perturbations, which we will label $B_0$ and $ T_0$ respectively, by combining
Eq.~(\ref{eq:poissalpha}) with Eq. (\ref{eq:bispPhi}) and (\ref{eq:trispPhi}). 

The fact that the Bardeen's potential decays with time proportionally to 
$g(z)$ implies
that the actual values of the coefficients $ Q_{\mathrm{NL}j}$ depend on the
cosmic epoch at which  Eq.~(\ref{eq:basichi}) is applied (see Section 2.2 in
\cite{Pillepich:2008ka}). Here we apply it at early times 
(which is sometimes called the ``CMB convention'') while
other authors use the fields linearly extrapolated at $z=0$ 
(the ``LSS convention''). In general,
\be
Q_{\mathrm{NL}j}^{\mathrm{LSS}}=Q_{\mathrm{NL}j}^{\mathrm{CMB}}
\left[\frac{g(\infty)}{g(0)} \right]^j
\ee
so that $\fnl^{\mathrm{LSS}}\simeq 1.3\, \fnl^{\mathrm{CMB}}$ and
$\gnl^{\mathrm{LSS}}\simeq 1.7\, \gnl^{\mathrm{CMB}}$. This conversion factors
should be taken into account when comparing papers using different conventions.

\section {Power spectra} \label{sec:pk}
We are now ready to calculate the two-point statistics of the LSS
arising from non-Gaussian initial conditions. 
We compute the halo-halo power spectrum and the halo-matter cross spectrum
as follows.
First,
we take the Fourier transform of Eq.~(\ref{eq:deltaE}) 
and build the corresponding
two-point correlators $\langle \tilde\delta_h(\mathbf k_1) \tilde 
\delta_h(\mathbf k_2)\rangle$
and $\langle \tilde\delta_h(\mathbf k_1) \tilde 
\delta(\mathbf k_2)\rangle$. 
These are composed of many pieces and
we only consider terms up to the fourth perturbative order.
For instance the leading contribution to halo-halo spectrum
(second order in terms of the perturbations) is composed of 3 terms, 
the third order correction 
is made of 8 pieces (of which one identically vanishes because $\varphi$
is a Gaussian field), and the fourth-order one
contains 30 terms (14 of which are obtained multiplying a linear perturbation
by a third-order one -- indicated by the subscript $_{(13)}$ hereafter -- 
and 16 are originated by
the product of two second-order terms -- subscript $_{(22)}$ hereafter).
To proceed we then:
(a) use Eq.~(\ref{eq:perturbn}) and write the density perturbations of order $n>1$ 
as convolutions of $n$ linear perturbations and a kernel $J_n$;
(b) express the linear density perturbations in terms of the potential $\Phi$
 using the Poisson Eq.~(\ref{eq:poissalpha});
(c) take the ensemble averages by using the expressions for the power
spectrum, bispectrum and trispectrum given in Eqs.~(\ref{eq:bispPhi}),
(\ref{eq:trispPhi}), (\ref{eq:deltaT}), and (\ref{eq:convspectra}).
While $\langle \tilde  \varphi (\mathbf k_1) \tilde  \varphi (\mathbf k_2) \tilde  \varphi (\mathbf k_3) \rangle=0$,
attention must be payed to the mixed terms in $\Phi$ and $\varphi$ as:
\be
\langle \tilde  \Phi (\mathbf k_1) \tilde  \Phi (\mathbf k_2) \tilde  \Phi
(\mathbf k_3) \rangle\simeq
\langle \tilde  \varphi (\mathbf k_1) \tilde  \Phi (\mathbf k_2) \tilde  
\Phi (\mathbf k_3) \rangle+
\mathrm{(2 \ cyc.)} 
\simeq
\langle \tilde  \varphi (\mathbf k_1) \tilde  \varphi (\mathbf k_2) \tilde  \Phi
(\mathbf k_3) \rangle 
+\mathrm{(2 \ cyc.)}
\propto \fnl
\ee
where equalities only hold at leading order in $\varphi^n$ 
(i.e. $\varphi^4$) as the 
central correlator also contains a sub-leading term proportional to $\fnl \gnl$
(which scales as $\varphi^6$) 
and the leftmost one some terms proportional to $\fnl^3$,
$\fnl \gnl$ (both scaling as $\varphi^6$), and $\fnl \gnl^2$ ($\propto \varphi^8$).
Similarly, to leading order in $\varphi^n$ (i.e. $\varphi^6$),
\ba
\langle \tilde  \Phi (\mathbf k_1) \tilde \Phi (\mathbf k_2) \tilde \Phi
(\mathbf k_3) \tilde \Phi (\mathbf k_4) \rangle&\simeq& \fnl^2 T_A+ 
\gnl T_B\\
\langle \tilde  \varphi (\mathbf k_1) \tilde \Phi (\mathbf k_2) \tilde \Phi
(\mathbf k_3) \tilde \Phi (\mathbf k_4) \rangle
+ \mathrm{(3 \ cyc.)}
&\simeq& 2 \,\fnl^2 T_A+
3 \,\gnl T_B\\
\langle \tilde  \varphi (\mathbf k_1) \tilde \varphi (\mathbf k_2) 
\tilde \Phi
(\mathbf k_3) \tilde \Phi (\mathbf k_4) \rangle+ \mathrm{(5 \ cyc.)}
&\simeq& \fnl^2 T_A+
 3\, \gnl T_B\\
\langle \tilde  \varphi (\mathbf k_1) \tilde \varphi (\mathbf k_2) 
\tilde \varphi
(\mathbf k_3) \tilde \Phi (\mathbf k_4) \rangle + \mathrm{(3 \ cyc.)}
&\simeq& \gnl T_B 
\ea
where $T_A(\mb k_1,\mb k_2,\mb k_3,\mb k_4)$ and
$ T_B(\mb k_1,\mb k_2,\mb k_3, \mb k_4)$ are defined in Eqs. (\ref{eq:trispPhi}) and (\ref{eq:deltaT}).

\subsection {Results and comparison with N-body simulations}

\begin {figure}
\begin{center}
\includegraphics[width=0.45\columnwidth,angle=0]{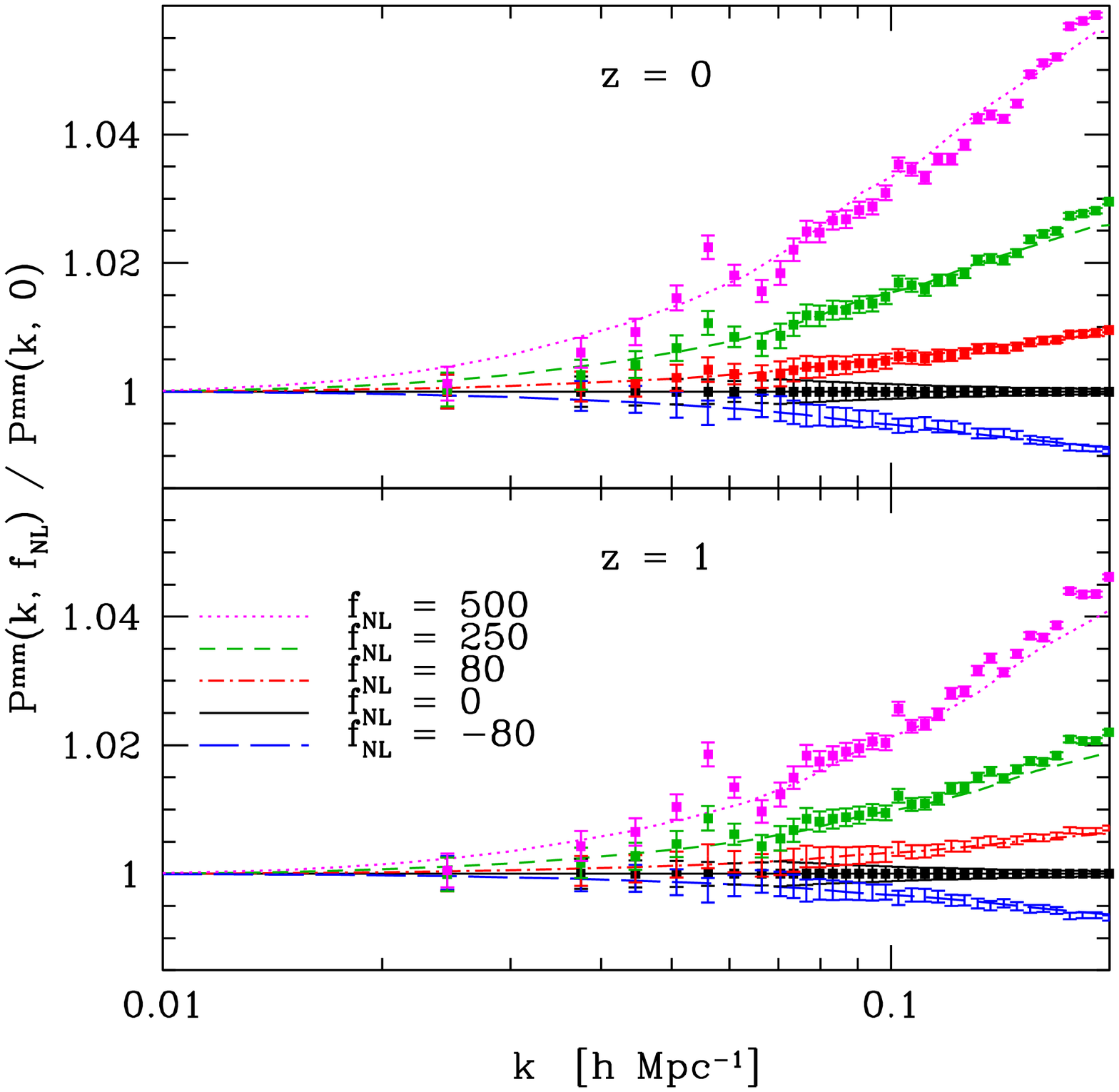}
\includegraphics[width=0.45\columnwidth,angle=0]{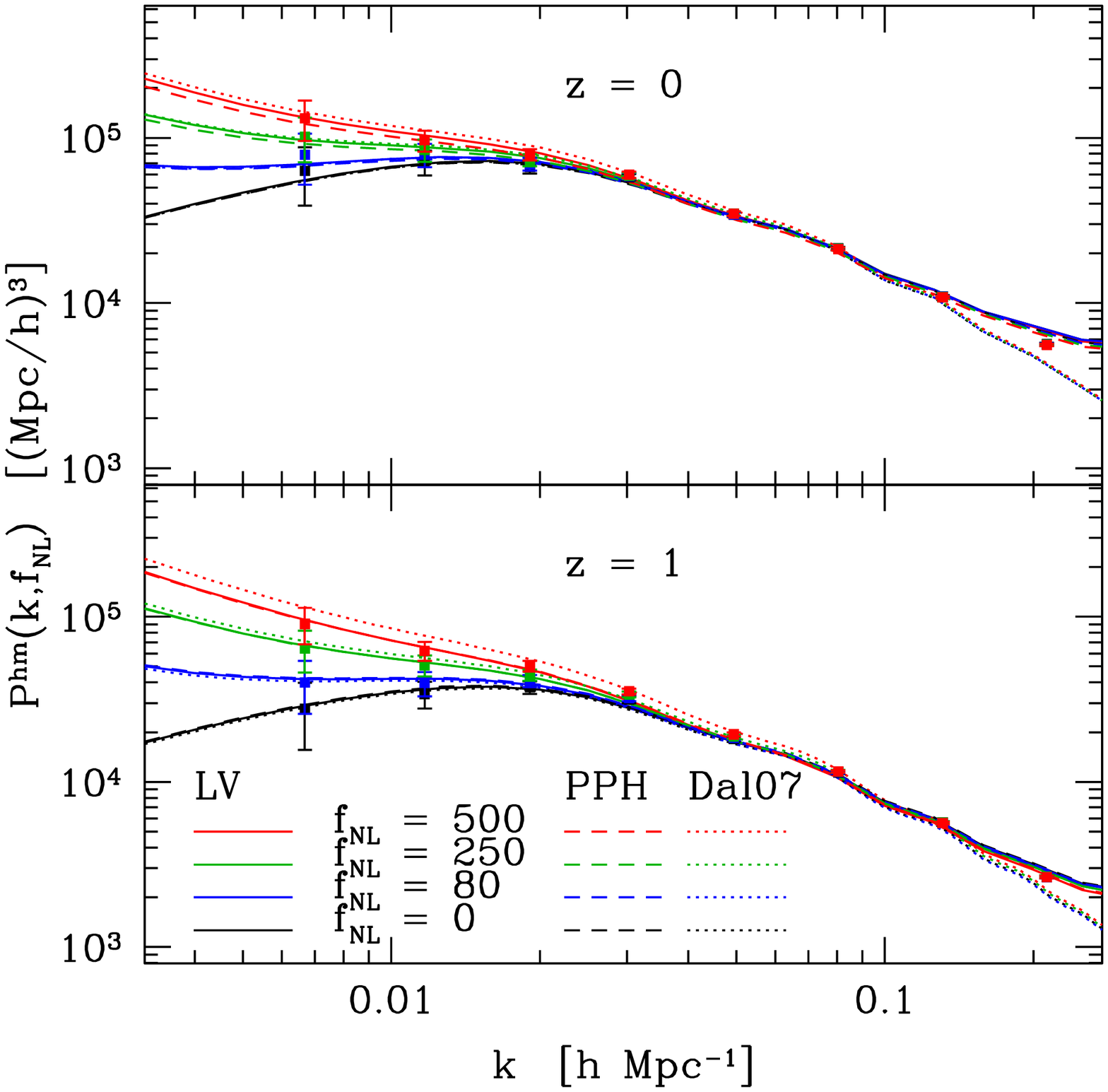}
\end{center}
\caption{\textbf{Left:} Deviation of the matter power spectrum 
in models with different $\fnl$ (and $\gnl=0$)
from the
Gaussian case at $z=0$ (top) and $z=1$ (bottom). The lines indicate our
one-loop calculation for different values of  $\fnl$ while
points with error bars correspond to the N-body simulations by PPH08. 
\textbf{Right:} Halo-matter cross spectrum 
at $z=0$ for a narrow bin of halo masses centered around 
$M = 2 \cdot 10^{14} M_{\odot}/h$ (top) and 
at $z=1$ for $M = 5 \cdot 10^{13} M_{\odot}/h$ (bottom).
The solid and dashed lines have been obtained using our model with
the bias parameters from the LV and PPH mass functions, respectively.
The dotted lines indicate the model by Dal07, while points with error bars
correspond to the simulations by PPH08.
}
\label{fig:Pmmk}  \label{fig:Phmk}
\end{figure}

\paragraph* {\bf Matter}
If we set $b_{10}=1$ and all the other bias coefficients to zero, 
we obtain an expression for the power spectrum of mass-density perturbations
which coincides with the result by \cite {Taruya:2008pg}:
$P^{mm}(k, z) = D^2(z) P_{11}(k) + D^3(z) P^{mm}_{12}(k) + D^4(z) \left[
  P^{mm}_{22}(k) + P^{mm}_{13}(k) \right] $, 
where
\ba
P^{mm}_{11} (k) &=& P_0 (k) \nn \\
P^{mm}_{12} (k) &=& 2 \int \frac{d^3 \mathbf q}{(2 \pi)^3} J^{(s)}_2 (\mathbf q, \mathbf k - \mathbf q) B_0 (- \mathbf k, \mathbf q, \mathbf k - \mathbf q) \nn \\
P^{mm}_{22} (k) &=& 2 \int \frac{d^3 \mathbf q}{(2 \pi)^3} \left[ J^{(s)}_2
  (\mathbf q, \mathbf k - \mathbf q) \right]^2  P_0(q) P_0(|\mb k - \mb q|) + \nn \\
&~& + \int \frac{d^3 \mathbf p \, d^3 \mathbf q}{(2 \pi)^6} \, J^{(s)}_2  (\mathbf p, \mathbf k - \mathbf p) \, J^{(s)}_2  (\mathbf q, -\mathbf k - \mathbf q) \,  T_0(\mathbf p, \mathbf k - \mathbf p, \mathbf q, -\mathbf k - \mathbf q) \nn 
\ea
\ba
P^{mm}_{13} (k) &=& 6 \int \frac{d^3 \mathbf q}{(2 \pi)^3} J^{(s)}_3 (\mb k, \mb q, -\mb q) P_0(q) P_0(k) +  \nn  \\
&~& + 2 \int \frac{d^3 \mathbf p \, d^3 \mathbf q}{(2 \pi)^6} \, J^{(s)}_3 (\mb p, \mb q, \mb k - \mb p -\mb q) \,  T_0(-\mb k, \mb p, \mb q, \mb k - \mb p -\mb q),
\ea
and $J_n^{(s)}$ indicates a kernel which has been symmetrized with respect
to its arguments.
We have checked that the two-loop contributions proportional to $T_0$ are, in
general, negligibly small and will not be considered hereafter if $\gnl=0$.
In the left panel of  Fig. \ref {fig:Pmmk} we plot the 
ratio between the matter power spectra originating from a non-Gaussian 
model (with $\fnl \neq 0$ and $\gnl=0$) and from Gaussian initial conditions. 
We consider several values of $\fnl$ and we compare our analytical results
with data from the N-body simulations by PPH08 at both redshift 0 and 1.
Primordial non-Gaussianity 
alters the matter power spectrum at the few percent level 
for $k<0.2 \ h$ Mpc$^{-1}$
and these deviations are remarkably well
reproduced
by the one-loop corrections. 

\begin {figure}
\begin{center}
\includegraphics[width=0.45\columnwidth,angle=0]{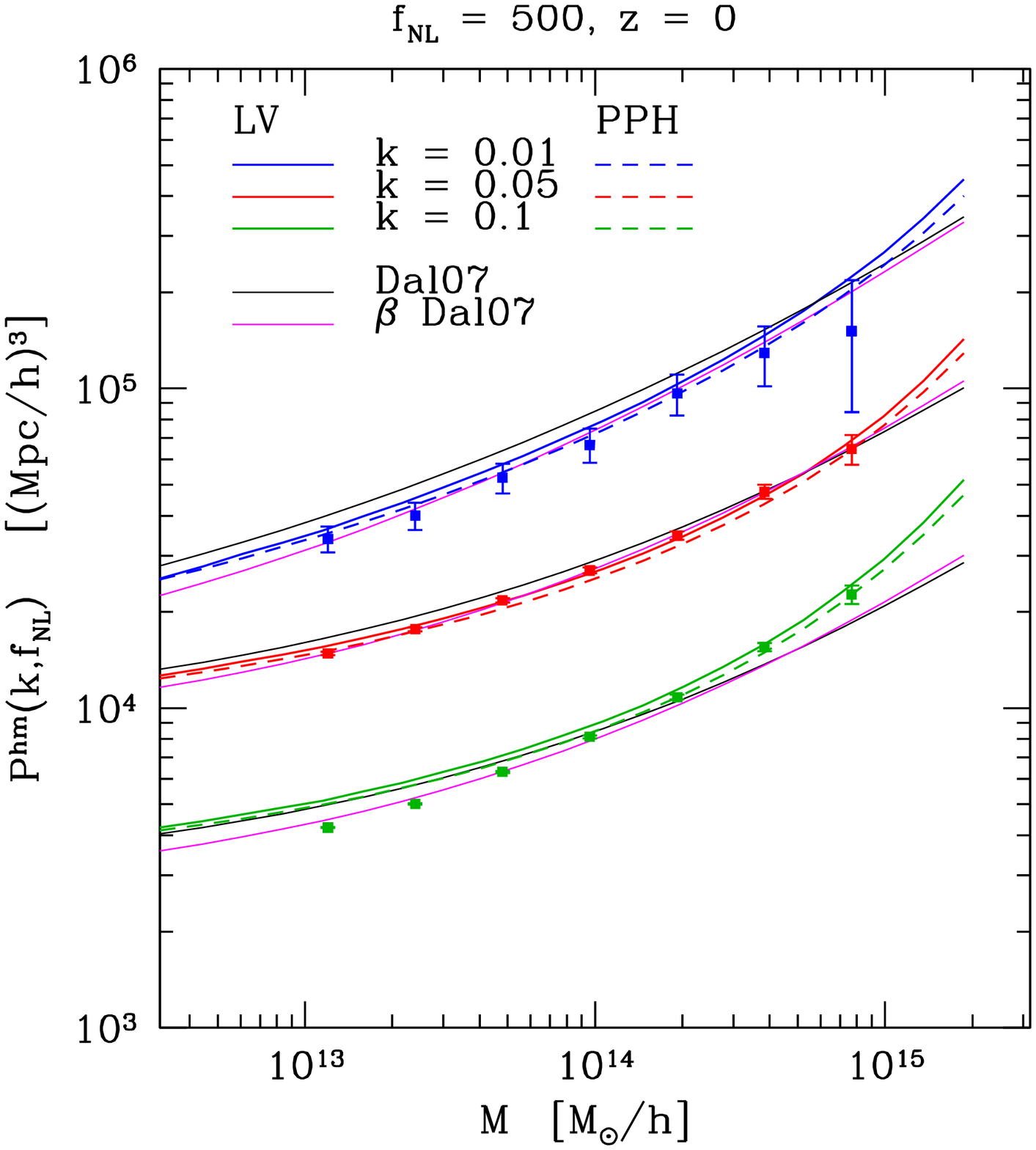}
\includegraphics[width=0.45\columnwidth,angle=0]{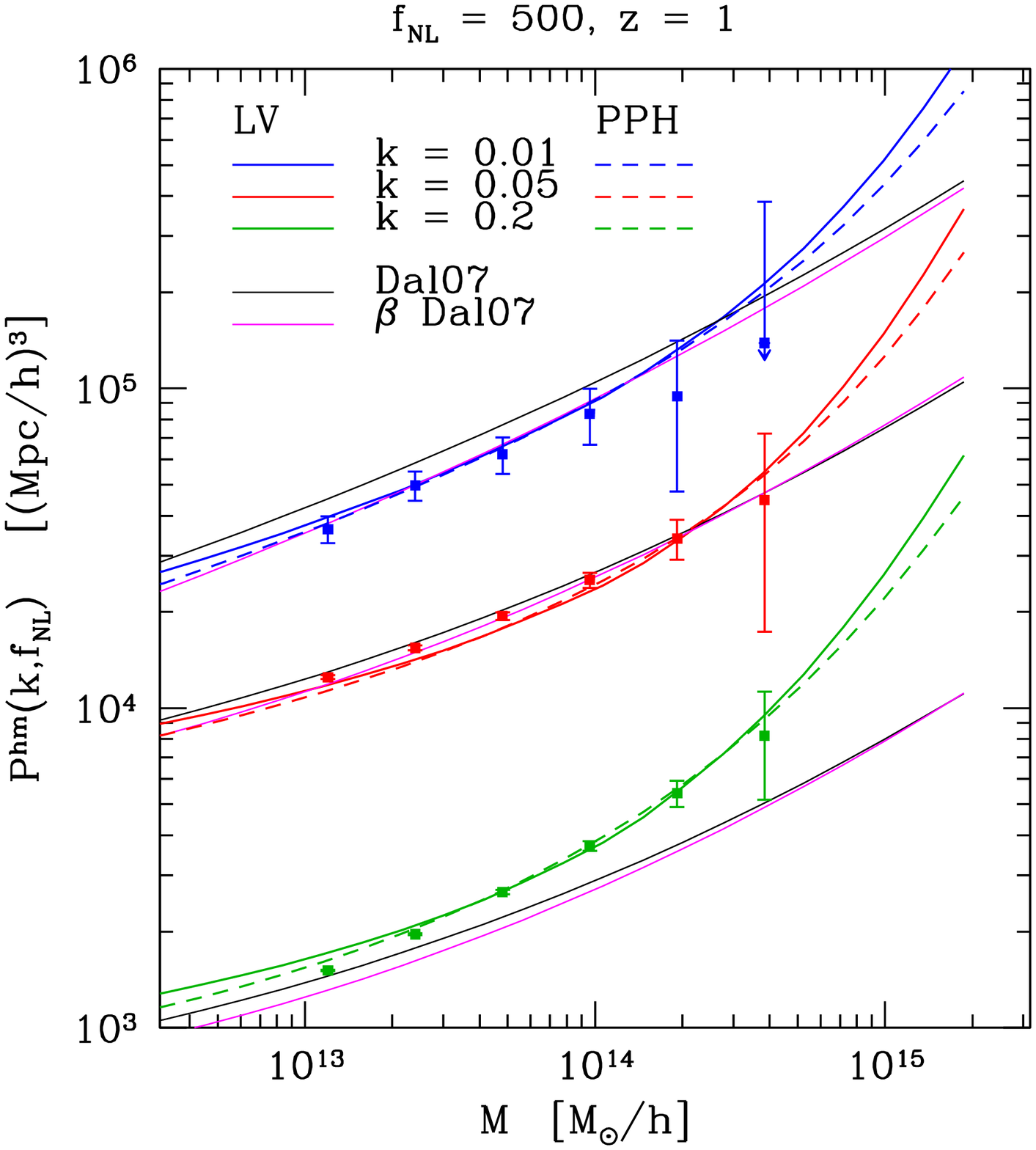}
\end{center}
\caption{The halo-matter cross spectrum as a function of halo
mass at $z=0$ (left) and $z=1$ (right) is plotted for three 
different values of the comoving wavenumber $k$ (in $h$ Mpc$^{-1}$).
Colored solid and dashed lines indicate the results of our perturbative calculation
using the LV and PPH mass functions, respectively.
Data points with error bars correspond to the N-body simulations by PPH08. 
The thin lines show the
linear theory by Dal07 (black), also corrected with the factor $\beta$ (magenta) introduced by PPH08 (see Section \ref{sec:asymp} for further details). 
 At $z=0$  structure has evolved further 
into the non-linear regime, so that the range of validity of both the linear 
and one loop theories is reduced. 
}
\label{fig:PhmM}
\end{figure}

\paragraph* {\bf Halos}
The halo-halo power spectrum 
(and similarly the halo-matter cross spectrum) 
deriving from our multivariate biasing scheme can be written as
\be
P^{hj}(k,z) =  D^2(z) \, P^{hj}_{11}(k) + D^3(z) \, P^{hj}_{12}(k) + D^4(z) \, \left[ P^{hj}_{22}(k) + P^{hj}_{13}(k) \right]\;,
\ee 
where the superscript $j$ indicates either matter ($m$) or halo ($h$) fluctuations.
The full expressions of the different terms are lengthy and we 
report them only in the Appendix. 
We highlight that our calculation reduces to: 
(a) the linear result by Dal07 if we only consider the leading-order terms
  (and further 
assume that the mass function does not depend on $\fnl$); 
(b) the usual one-loop Gaussian expression  
 derived e.g. by \cite {Heavens:1998es} if we set $\fnl=0$, and 
(c) the non-Gaussian result by \cite{Taruya:2008pg} if we ignore the terms 
which are proportional to the potential perturbations $\varphi$ in our
multivariate biasing scheme.
Notice that for $k \rightarrow 0$ we obtain $P^{hh}(k) \propto P_0(k)/\alpha^2(k)$ and $P^{hm}(k) \propto P_0(k)/\alpha(k)$.
In the right panel of Fig.~\ref{fig:Phmk}, we test our
theoretical predictions for the halo-matter cross spectrum as a function
of $\fnl$ and for $\gnl=0$ (solid lines) against the N-body data by PPH08.
The bias factors in the models 
have been calculated from the LV and PPH mass functions.
We consider halos with mass $M \simeq 2 \cdot 10^{14} M_{\odot}/h$ at $z=0$ and
$M \simeq 5 \cdot 10^{13} M_{\odot}/h$ at $z=1$. We have chosen two different
mass bins to keep the number of halos at each redshift large enough to avoid
substantial shot noise contamination in the simulations. 
Our analytical results are in very good agreement with the simulation data 
for the whole range of $\fnl$ and up to scales $k \lesssim 0.2 \, h$~Mpc$^{-1}$.
Note that our spectra differ from 
the linear result by Dal07 (over-plotted with dotted lines)
both on large and small scales.
The small-scale departure is due to the non-linear growth of perturbations
which we take into account up to the third perturbative order.
The large-scale discrepancy is discussed in detail in the next subsection.
In Fig.~\ref{fig:PhmM} we show how the cross spectrum depends on the 
halo mass at three different wavenumbers and for two redshifts.
Independently of halo mass, 
at $z=1$ our model accurately matches the outcome of the simulations for 
$k<0.2 \, h$~Mpc$^{-1}$. Likewise, at $z=0$, the theory agrees well with
the numerical data on the largest scales  while
it tends to over-predict
the cross power for $k>0.1 \, h$~Mpc$^{-1}$ and $M<10^{14} M_\odot/h$.

We have checked that, for $\gnl = 0$, the terms proportional to the trispectra of the
potentials $\Phi$ and $\varphi$ are generally subdominant even when
they generate contributions to the halo power spectrum
which diverge as $k\to 0$.
For instance, the trispectrum contribution to the term
$\langle \widetilde{\delta_1^2} \widetilde{\delta_1^2} \rangle$ in the
halo-halo power spectrum scales as $P_0(k)/\alpha^2(k)$ as like as the
leading term. However, for $|\fnl|<500$, this correction contributes 
at most at percent level and only on very large scales.
Also the trispectrum contribution 
in $ \langle \widetilde{\delta_1^3} \tilde\delta_1 \rangle$ which
scales as $P_0(k)/\alpha(k)$ is subdominant ($\ll 1\%$)
for both the halo-halo and the halo-matter cases.
Similar conclusions can be drawn for the terms including $\varphi$:
we have checked that the contributions arising from averages where one or more 
$\delta_1$ are replaced by $\varphi$ are subdominant.

\paragraph* {\bf The effect of $\mathbf {\gnl}$} The situation becomes more complicated 
if we consider also the third-order term in Eq.~(\ref{eq:basichi})
with realistic values of $\gnl$. In this case, there will be new contributions to the halo and halo-matter power spectra coming from two sources: 
 the trispectrum gets the additional term of Eq.~(\ref{eq:deltaT}), and the bias factors become altered as described in Eq.~(\ref{eq:dbgnl}); each of these modifications is linear in $\gnl$.

Considering first the effect of $\Delta T_{\Phi}$, we have found that this adds a (negligible) constant contribution to 
$\langle \widetilde{\delta_1^2} \widetilde{\delta_1^2} \rangle$ 
but generates another term which scales as  $P_0(k)/\alpha(k)$
in $\langle \tilde\delta_1  \widetilde{\delta_1^3} \rangle$.
The latter
can become the dominant contribution 
on large scales and for high values of $\gnl \gtrsim 10^5$. 
In the limit $k \to 0$, this two-loop term (whose full expression is given in Eqs.~(\ref{eq:PII13},\ref{eq:PII13b}) in the Appendix for the halo and halo-matter cases) reduces to
\ba \label{eq:triII}
P_{13}^{hh, II} (k) &\to& \gnl \, b_{10} b_{30} \, \sigma^4 (R) \, \Sigma_3(R) \, \frac{P_0(k)}{\alpha(k)} \propto \gnl \, k^{n_s-2} \, \nn \\
P_{13}^{hm, II} (k) &\to&  \frac 1 2 \,\gnl \, b_{30} \, \sigma^4 (R) \, \Sigma_3(R) \, \frac{P_0(k)}{\alpha(k)} \propto \gnl \, k^{n_s-2} \, ,
\ea
where we define
\be
\Sigma_3(R) \equiv \int \frac{d^3 \mb p \, d^3 \mb q}{(2 \pi)^6} \, \frac{P_0(q)}{\alpha(q)}\, \frac{P_0(p)}{\alpha(p)} \, \alpha (|\mb p + \mb q|) \, W(q R) \, W(p R) \, ,
\ee
such that $S_3 = \fnl \, \Sigma_3$.
In our formalism, $P_{13}^{hh, II} (k)$
corresponds to the leading 
contribution 
 found by \cite {Desjacques:2009jb},
who used it to derive observational constraints on $\gnl$ (assuming
$\fnl=0$). In the limit of high peaks, the scaling $b_{10} b_{30} \to (\delta_c / \sigma)^4$ is recovered.
Note, however, that the amplitude of this term depends on
the adopted smoothing scale $R$ first introduced in Section~\ref{sec:tracers}, and further discussed in Section~\ref{sec:smoo}.

 Second, we look at the effect of $\Delta b$. As shown in Eq.~(\ref{eq:dbgnl}), only the coefficients $b_{02}$ and $ b_{12}$ are altered by $\gnl$. In the halo-matter case these new terms are subdominant with respect to the trispectrum contribution of Eq.~(\ref{eq:triII}). However, in the halo-halo spectrum, the leading term for $k \to 0$ is $P^{hh}_{(23)(23)} \propto \langle \widetilde {\varphi^2} \widetilde {\varphi^2} \rangle \propto b_{02}^2$, which scales as
\be
P^{hh}_{(23)(23)} \to \frac 1 2 \, b_{02}^2 \, \sigma^2_{\varphi} (R) \, \frac{P_0(k)}{\alpha^2(k)} \propto  \gnl^2 \, k^{n_s-4} \, ,
\ee
where $\sigma^2_{\varphi} = \int d q \, q^2 \, [P_0 (q) / \alpha^2 (q)] \, W^2 (q R) / (2 \pi^2)$.
Because of the quadratic dependence on $\gnl$, this term dominates on very large scales for high values of $\gnl$ and small $\fnl$. Its dependence on the smoothing radius $R$ is very weak because the potential $\varphi$ is nearly scale invariant.
Note that the terms in $\fnl \gnl$ are generally subdominant.

\subsection{Bias and asymptotic behavior on large scales} \label{sec:asymp}
\begin {figure}
\begin{center}
\includegraphics[width=0.45\columnwidth,angle=0]{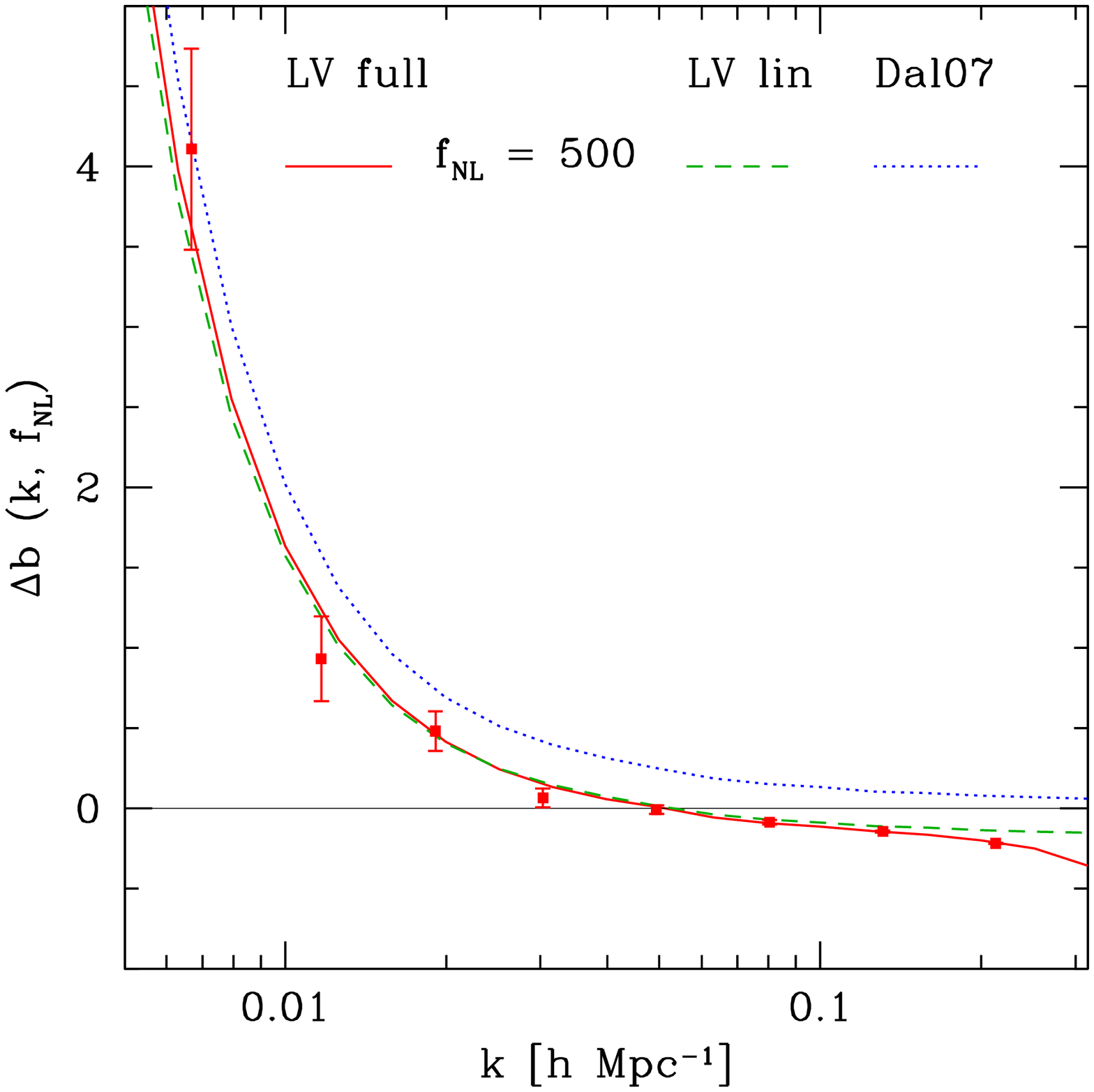}
\includegraphics[width=0.45\columnwidth,angle=0]{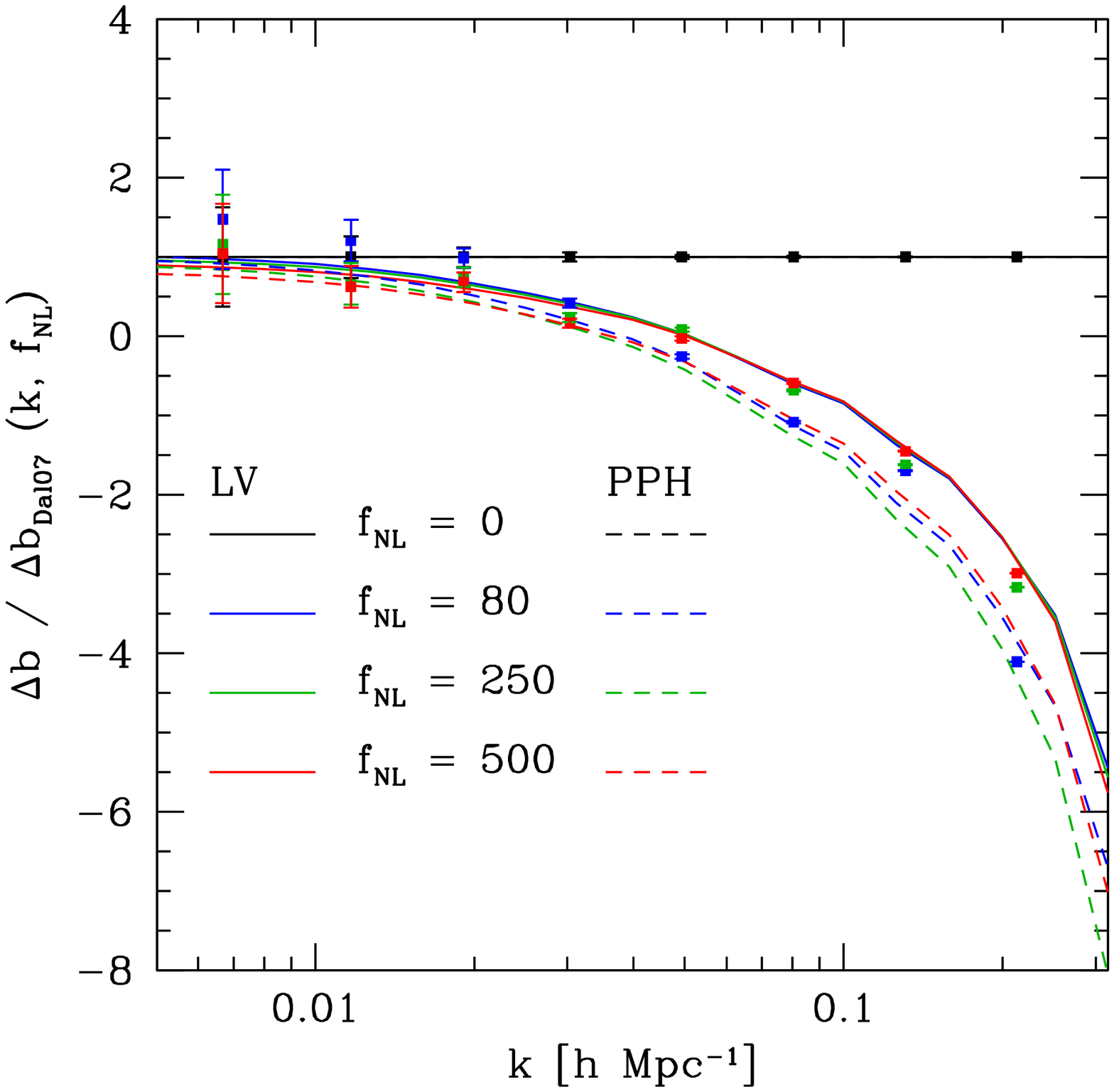}
\end{center}
\caption{  
\textbf{Left:} 
Change in the effective bias $\Delta b$  for fixed $\fnl=500$, $M = 2 \cdot 10^{14} M_{\odot}/h$, at $z=0$.  Different models are compared with N-body simulations: 
the simple $\Delta b_{\mathrm{Dal07}}$ (dotted),
$\Delta b_{\mathrm{linear}}$ (dashed) using the LV mass function,
and the full one-loop theory (solid), which yields the best match.
\textbf{Right:} Fractional difference between the one-loop prediction for $\Delta b$ (with LV and PPH mass functions)
  and the Dal07 linear theory, compared with the simulations, for different values of $\fnl$, at the same mass and redshift.
 }
\label{fig:db} \label{fig:dbratio}
\end{figure}

Let us define the effective bias function
\be
b_{\mathrm{eff}} (k, \fnl) \equiv \frac {P^{hm} (k, \fnl)}{P^{mm} (k, \fnl)}\;,
\ee
and compare it with the standard Gaussian bias by introducing the bias deviation
\be
\Delta b (k, \fnl) = b_{\mathrm{eff}} (k, \fnl) - b_{\mathrm{eff}} (k, 0).
\ee
In this section we will only consider the case $\gnl =0$. 
In the limit $k \to 0$ and for large $R$ ($\sigma_R^2 \ll 1$), 
the dominant contribution to the halo power spectrum
is given by the tree-level term, and we thus obtain
\be 
\label{eq:dbaaa}
\Delta b_{\mathrm{linear}} (k) = b_{10} (\fnl) - b_{10} (\fnl = 0) + 2\fnl \delta_c \, [b_{10}(\fnl)-1]/\alpha(k)\;.
\ee
The scale-dependent non-Gaussian correction is proportional to the factor
$b_{10}-1$ as originally shown by Dal07, although there is also an additional 
scale-independent correction, due to the fact that in our model $b_{10}$ is a 
function of $\fnl$ (similar conclusions have been reached by \cite{Slosar:2008hx,
  Desjacques:2008vf, Pillepich:2008ka, Desjacques:2009jb, Valageas:2009vn} following different
approaches). 
In the simplest model by Dal07 the scale-independent term 
is missing:
\be
\Delta b_{\mathrm{Dal07}} (k) =  2\fnl \delta_c \, [b_{10}(\fnl)-1]/\alpha(k)\;.
\label{eq:dalal}
\ee
In the left panel of
Fig.~\ref{fig:db}, we plot $\Delta b (k)$ for $\fnl=500$ and 
test the different models against the N-body simulations by
PPH08. 
We can see that considering only the scale-dependent term as in Dal07 does not
match the simulations very well, since, contrary to the N-body data, 
$\Delta b$ cannot change sign with increasing $k$ (see also  \cite{Pillepich:2008ka}).
The agreement vastly improves if we use Eq.~(\ref{eq:dbaaa}) with the bias 
parameters  computed from a non-Gaussian mass function (LV) as 
this adds a constant negative shift (see the left 
panel of Fig.~\ref{fig:b1b2b3}) 
to the bias deviation.
Considering the full calculation to third perturbative order further improves
the agreement with the simulations for $k>0.1\ h$ Mpc$^{-1}$ up to 
a maximum value of the wavenumber 
which depends on the adopted smoothing scale for the
perturbative calculations (see Section~\ref{sec:smoo} for further details).
\begin {figure}
\begin{center}
\includegraphics[width=0.45\columnwidth,angle=0]{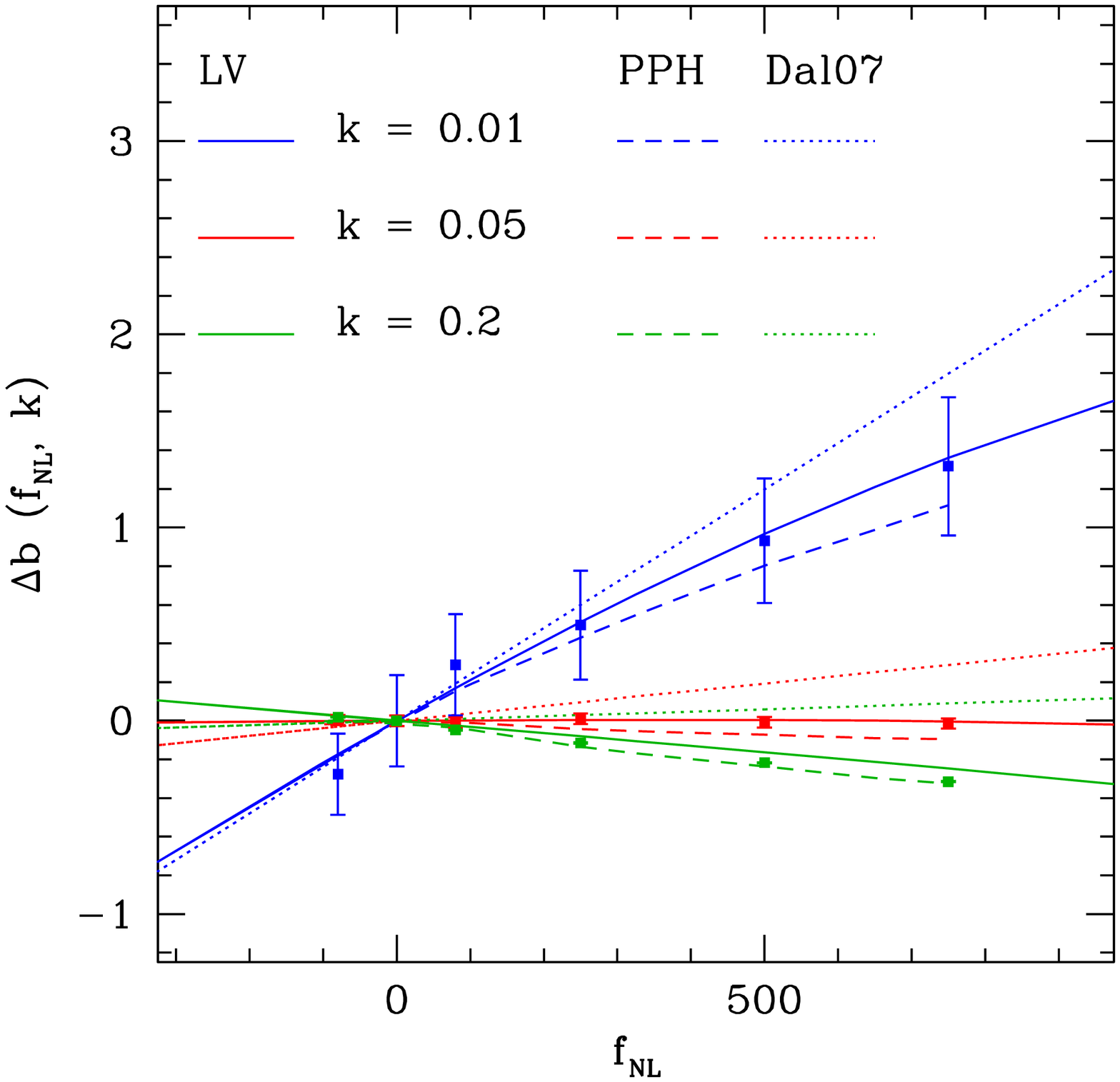}
\includegraphics[width=0.45\columnwidth,angle=0]{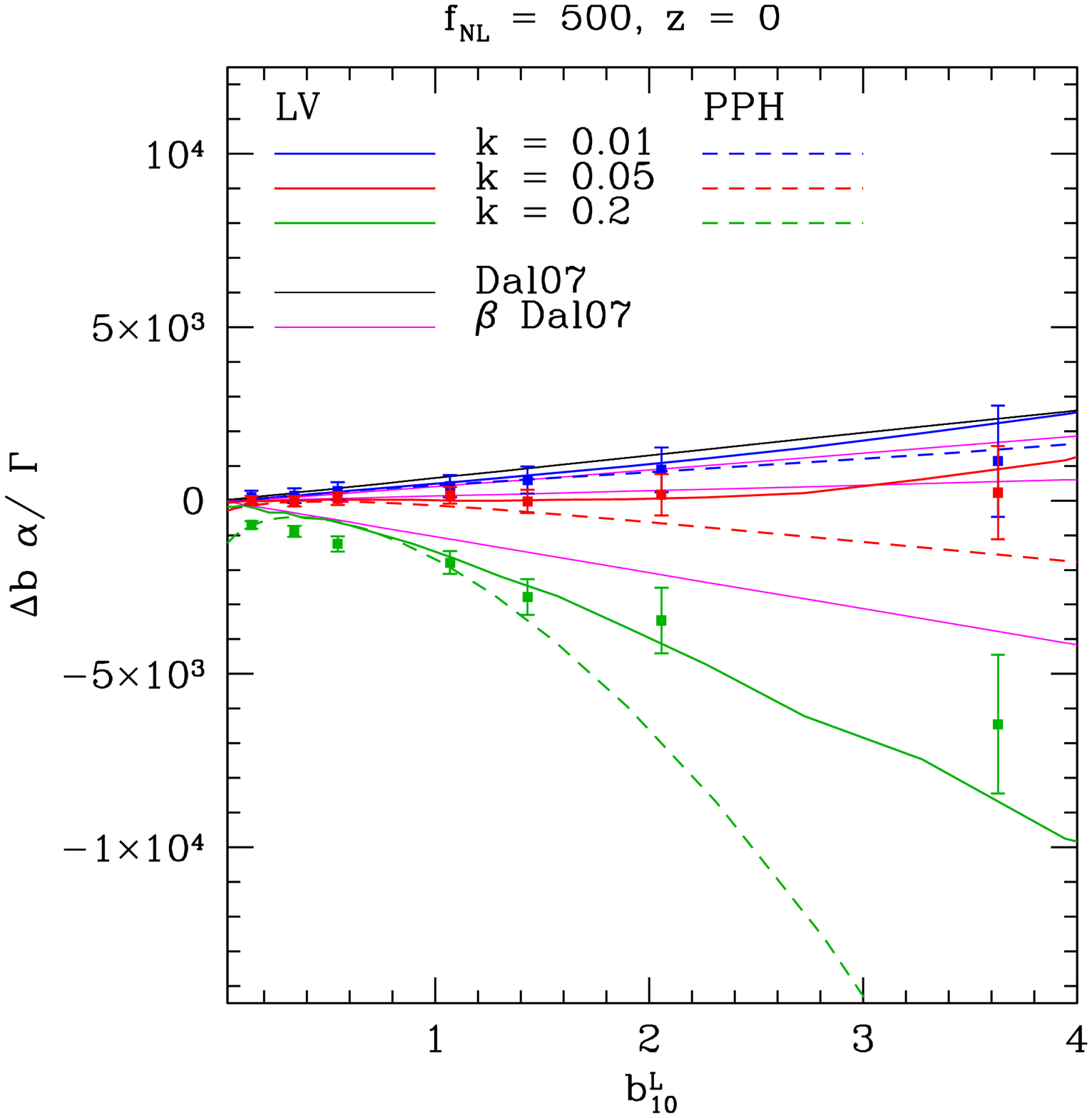}
\end{center}
\caption{\textbf{Left:} Bias deviation at $z=0$ as a
function of $\fnl $ at three different wavenumbers, and for
  $ M = 2 \cdot 10^{14} M_{\odot}/h$. The results for LV (solid) and PPH (dashed, within its range of validity) mass functions are shown.  We also plot the prediction for the linear (Dal07) theory.
\textbf{Right:} As in the left panel but as a function of $b_1^L $ and for $\fnl = 500$. The normalization factor $\alpha/\Gamma \equiv \alpha / (2 A)$ is chosen to reproduce figure 11 in PPH08.
}
\label{fig:dbfnl} \label{fig:dbb}
\end{figure}
To highlight the importance of the non-linear and scale-independent
corrections,
in the right panel of Fig.~\ref{fig:dbratio} we
plot the ratio 
\be
\frac {\Delta b (k, \fnl)} {\Delta b_{\mathrm{Dal07}} (k, \fnl)}
\ee
for several values of $\fnl$ and using both the 
LV and PPH mass functions. We consider halos
with mass $M = 2 \cdot 10^{14} h^{-1} M_{\odot}$ at $z=0$. 
The simulation data by PPH08 
are in good agreement with our third-order calculation, 
while the linear (Dal07) model cannot reproduce them 
on scales $k > 0.02 \, h$~Mpc$^{-1}$.
The impact of primordial non-Gaussianity on the halo bias 
is further explored in the left panel of
Fig.~\ref{fig:dbfnl}, where we show the bias deviation as a function of 
$\fnl$ at three selected scales, again for halos with 
$M = 2 \cdot 10^{14} h^{-1} M_{\odot}$ at $z=0$. 
Note that, contrary to what predicted by the linear (Dal07) model,
the relationship between $\Delta b$ and $\fnl$ is non-linear, 
as first observed by PPH08 in their N-body simulations, 
and this is now fully explained by our perturbative 
calculation to third order.
We finally study the mass dependence of the bias deviation 
by showing, in the right panel of Fig.~\ref{fig:dbb}, how $\Delta b$ 
changes as a function of the first Lagrangian bias coefficient $b_{10}^L$.
We consider three values of the wavenumber in the 
quasi-linear and mildly non-linear regime, $\fnl =500 $, and $z=0$.  
In the Dal07 model, $\Delta b$ depends linearly on $b_1^L$ and 
the relation between these two quantities is independent of $k$.
This is indicated by the solid black line which 
departs more and more from the simulation data with increasing $k$.
In order to better describe the N-body results,
PPH08 presented a fitting function, $\beta(k,\fnl)$, 
in the form of a multiplicative
(scale-dependent) correction to the Dal07 model (magenta lines).
Note that our third-order calculations match well the numerical outcome.
Some discrepancy is noticeable for $k=0.2\ h$ Mpc$^{-1}$ and $b_{10}^L<1$,
where non-linear effects become more important (due to the small first
bias coefficient) and perturbation theory becomes less accurate. 
The difference between the bias deviations obtained with the LV and PPH mass
functions at large halo masses emphasize the need for accurate parameterizations
of the halo counts for the rarest objects.

Our result in Eq.~(\ref{eq:dbaaa}) differs from 
the perturbative calculations based on the univariate local bias by
\citep{Taruya:2008pg, Sefusatti:2009qh}
where the scale-dependent part of the bias deviation was found to
scale as $b_{20}$ times the variance of the mass density field.
Strictly speaking, for $\fnl \neq 0$ and $k\to 0$ our result gives 
$P^{hm}(k,\fnl) \to (b_{01}+2 b_{20} \fnl \sigma^2_R) P_0(k)/\alpha(k)$
but the term proportional to $b_{20}$ is suppressed by smoothing
on the scale $R$ (which is necessary to truncate the bias expansion 
at third order in a meaningful way).
Note that, using the PS expression for the bias parameters in
the limits of high peaks, $\delta_c/\sigma \gg 1$, this reduces to
$P^{hm}(k,\fnl) \to 2 \fnl (\delta_c^2/\sigma^2) [1+ (\sigma^2_R/\sigma^2)]P_0(k)/\alpha(k)$. 
In this case, 
the two contributions are identical if $R$ is chosen to be the
Lagrangian radius of the halos (i.e. $R=R_f$) 
and the same window function is used
to compute $\sigma$ and in the calculation of the perturbative power spectra.
In general, however, using the Lagrangian radius of the halos gives
$\sigma$ of order unity and this is too large to allow the truncation of the
bias expansion at a finite order. For this reason in our calculations we
use $\sigma_R<\sigma$ and the smoothing-dependent 
contribution proportional to $b_{20}$ is subdominant.\footnote{
It is interesting to see what alternative approaches find regarding this
discrepancy.
For instance, Matarrese \& Verde \cite{Matarrese:2008nc} computed the
two-point correlation function
of regions where the density exceeds a high threshold $\delta_c\gg \sigma$ and
found that, for large separations,
the non-Gaussian correction scales as
 $(\delta_c^3/\sigma^6)\, \xi_3(\mb x_1,\mb x_1, \mb x_2)$ with $\xi_3$ the
three-point correlation function of the mass density. In Fourier space
this coincides with the high-peak limit of our result but, provided that
$\sigma^2=\sigma^2_R$, it also matches the result by \citep{Taruya:2008pg}.
This happens because for high peaks both 
$\delta_c (b_{10}^L)^2$ and $b_{10}^L b_{20}^L$ are proportional to $\delta_c^3$.
The same ambiguity
applies to the higher-order calculations in
Desjacques \& Seljak \cite{Desjacques:2009jb}.}

\begin {figure}
\begin{center}
\includegraphics[width=0.45\columnwidth,angle=0]{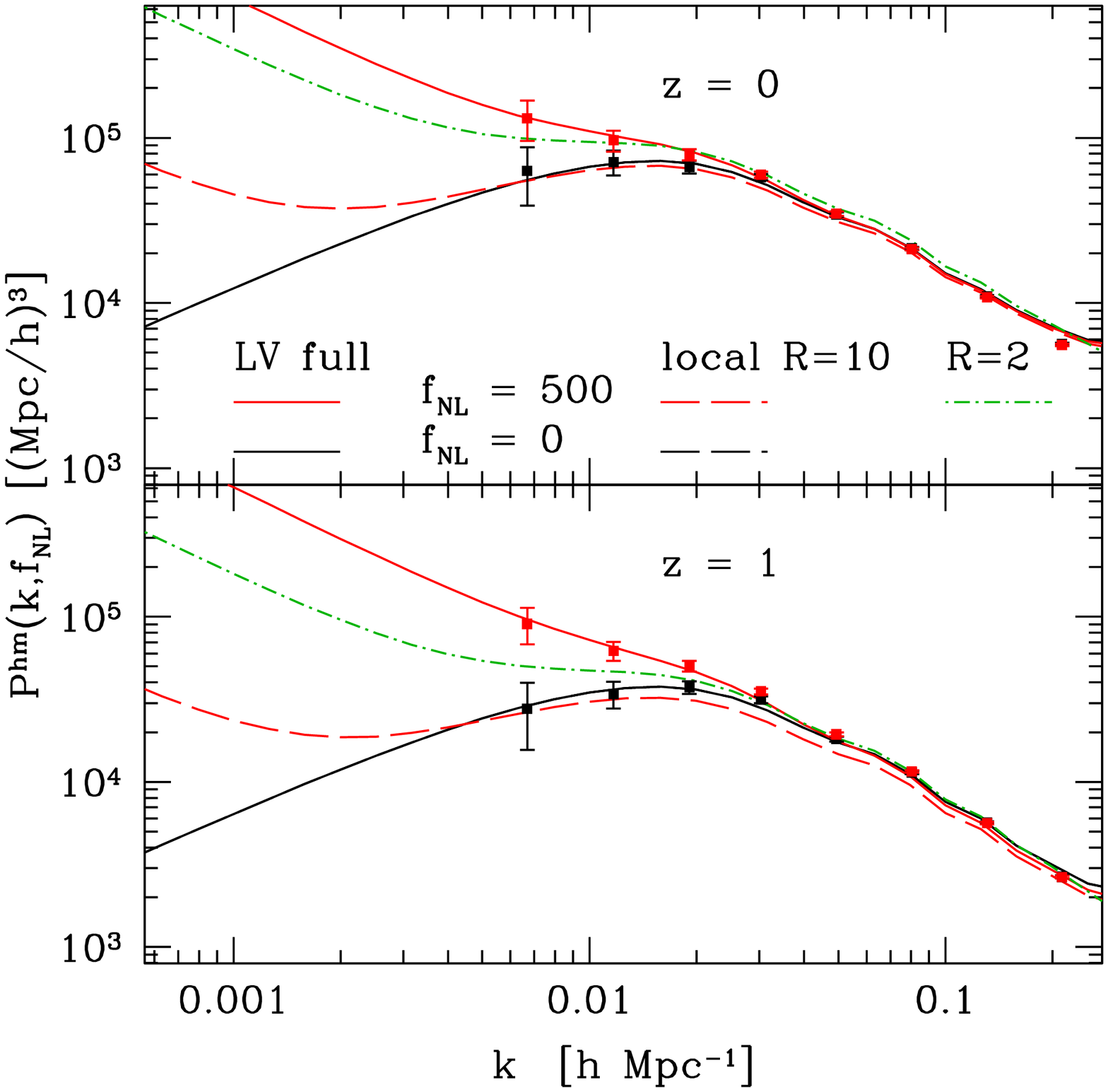}
\includegraphics[width=0.45\columnwidth,angle=0]{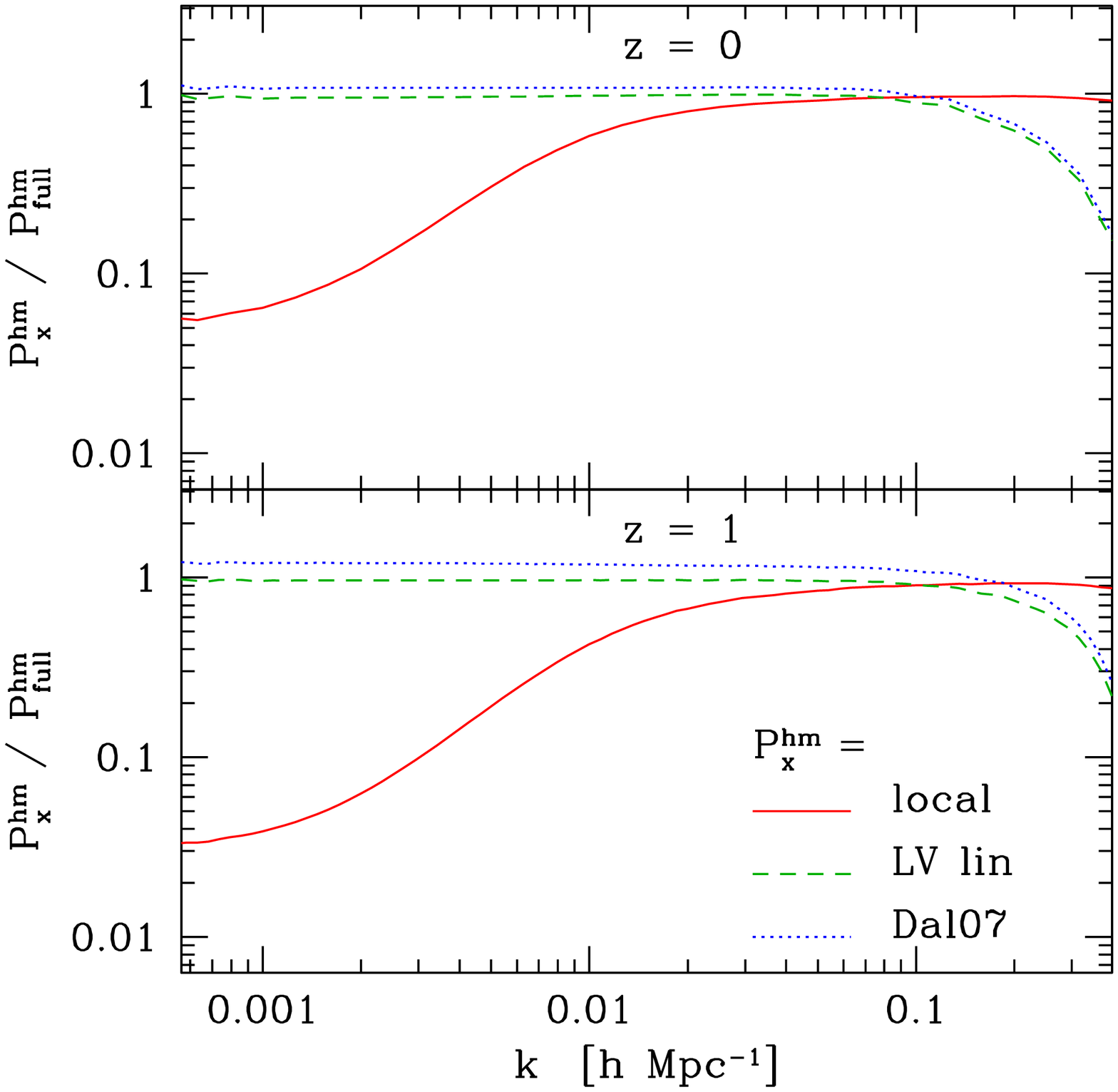}
\end{center}
\caption{ \textbf{Left:} The standard univariate local bias 
prescription (local) is compared 
with our multivariate scheme for the halo bias (LV full).
We consider the same halo masses as in Fig.~\ref{fig:Phmk} (right) and we 
plot the halo-matter cross spectra deriving from the two bias models 
at $z=0,1$ and with a smoothing scale $R=10 h^{-1}$ Mpc. In the Gaussian case ($\fnl=0$) the cross spectra coincide,
but they are very different for $\fnl=500$. 
While the asymptotic scale dependence for $k \rightarrow 0$ is recovered in both cases, only the multivariate model can reproduce the results from the N-body simulations. We also show  the dependence on $R$ in the univariate model, by overplotting the result with  $R=2 h^{-1}$ Mpc.
\textbf{Right:} Ratio between the halo-matter cross spectra 
obtained with different approximations 
and our full perturbative calculation, using $\fnl=500$ 
and for the same redshifts and halo masses as in Fig~\ref{fig:Phmk}. 
Assuming 
univariate local biasing (red, solid) severely under-predicts the 
large-scale power. 
On the other hand, the linear approximations in Eqs.~(\ref{eq:dbaaa}) 
(green, dashed) and (\ref{eq:dalal}) (blue, dotted) lack small-scale
power. Notice that 
the Dal07 model also features a constant offset on large scales
due to the missing scale-independent correction discussed in the main text. 
}
\label{fig:important}
\label{fig:importantrelat}
\end{figure}

As highlighted in Eq.~(\ref{eq:bfexpEuler}) the leading order for $\delta_h$ in
our multivariate biasing scheme includes a term proportional to $\varphi$ and
this generates the scale-dependent correction in $\Delta b$. 
The proportionality with $b_{20}$ found by other authors derives from 
the assumption
that a local deterministic bias scheme holds true
also in the presence of non-Gaussian perturbations. In this case, for
$k\to 0$, second-order terms dominate over
the tree-level contribution to the power spectrum which casts some doubts on
the validity of the perturbative expansion.
In the left panel of Fig.~\ref{fig:important} we show the difference between our 
multivariate approach and the standard local bias.
The asymptotic scale dependence $ P^{hm}(k) \propto \alpha^{-1}(k) \, 
P_0(k) \propto k^{n_s-2} $ for $k \rightarrow 0$ is recovered in both cases,
but the amplitude of the diverging term 
in the standard local bias model depends on the smoothing length
that has to be introduced to cure the ultraviolet divergence of the mass
variance. In  Fig.~\ref{fig:important} we use a smoothing scale of 10 $h^{-1}$ Mpc 
for both models and the asymptotic term deriving from the standard local bias
is strongly subdominant with respect to the correction given in 
Eq.~(\ref{eq:dbaaa}). As discussed by \cite{Taruya:2008pg}, the result of the univariate local model depends strongly on the smoothing scale, as it is proportional to $\sigma^2(R)$.
For instance, using $R = 2 h^{-1}$Mpc boosts the amplitude of the scale-dependent bias (see  Fig.~\ref{fig:important} (left)). 
This is why only the multivariate model can 
reproduce the results from N-body simulations by PPH08 without tuning additional parameters.
The main practical advantage in this case 
is that the bias parameters can be predicted from a model
for the mass function, while the results from 
the standard local bias can only be used after ``renormalizing''
the bias coefficients \cite{McDonald:2006mx} and using them as fitting functions.
However, even though one can play with the parameters of the theory to fit
some data, one should not forget that 
the physical origin of the scale-dependent bias is that large-scale 
fluctuations in $\delta_h$ trace
perturbations in $\varphi$ and this is not accounted for by the standard local
bias model.
The differences between the models are further highlighted in the right panel of
Fig.~\ref{fig:importantrelat}, 
where we plot the ratio of the halo-matter cross spectra obtained with 
different approximations with respect to our full non-Gaussian one-loop 
calculation, at $\fnl=500$ and at $z=0$ and $1$. 
This figure summarizes all the conclusion we have reached in
this section: (1) the univariate local biasing assumption yields the correct
$k$-dependence but the wrong amplitude of the spectrum on large scales;
(2) the linear approximation in Eq.~(\ref{eq:dbaaa}) lacks small-scale power;
(3) the simpler model by Dal07 also features a scale-independent offset in
the effective bias.
Similar results can be obtained for the halo-halo power spectrum, 
for which the asymptotic scale dependence is $ P^{hh}(k) \propto \alpha^{-2}(k) \, P_0(k) \propto k^{n_s-4} $ for $k \rightarrow 0$ .

\subsection {On the smoothing and other perturbative approaches} \label{sec:smoo}

The series expansion in Eq.~(\ref{ldb}) only applies 
when $\delta_h$ and $\delta$ have been smoothed on a scale $R$.
The reason is twofold. First, the locality of the bias is expected to 
degrade progressively when smaller and smaller scales are considered. 
Second, we truncate both the bias and the perturbative expansions
to finite order which is a good
approximation only if the neglected terms give a small contribution.
This requires that typically $\delta \ll 1$, i.e. $R\gg 5 h^{-1}$ Mpc.

As explicitely indicated in the Appendix, in this paper
we have used a Gaussian kernel $W(kR)=\exp[-(kR)^2/2]$
with $R=10\ h^{-1}$ Mpc to smooth the evolved fields $\delta_1$, $\delta_2$ 
and $\delta_3$ before applying Eq.~(\ref{ldb}).
An obvious consequence of this procedure is that also the resulting
halo power spectrum is suppressed on scales $ k \gtrsim R^{-1}$. 
A common prescription to lessen this effect and extend 
the theory to (slightly) larger wavenumbers is to divide 
out $W^2(kR)$ from the perturbative result for $P^{hh}(k)$ and $P^{hm}(k)$, 
as introduced by \cite {Smith:2006ne}. Here we have followed this approach
to consider wavenumbers up to $k\sim 0.3 \ h$ Mpc$^{-1}$.

Some of 
the one-loop corrections to the halo power spectrum present ultraviolet
divergences that are automatically cured by using a finite value of $R$.
However, some of these integrals give rise to scale-independent contributions
for $k\to 0$ whose amplitude depends on $R$,
as shown already by \cite {Heavens:1998es}. 
This is somewhat unsatisfactory, since it makes the 
results dependent upon a non-fundamental quantity, and it is amongst the 
reasons which have led to the application of renormalization techniques to 
the theory, often borrowed from other areas of physics.
The existing approaches, as recently reviewed by \cite {Carlson:2009it}, 
include the renormalized perturbation theory 
\cite{Crocce:2005xy,Crocce:2005xz}, 
the closure theory \cite{Taruya:2007xy}, 
the time renormalization group flow model \cite{Pietroni:2008jx}, 
and the renormalization group perturbation theory 
\cite{McDonald:2006hf,McDonald:2006mx}.
Most of these approaches do not include a bias model and just
apply to the matter density field. 
The renormalization of the bias parameters, 
included in some of the models, makes the theory free from any undesired 
dependence on the smoothing scale. 
This can be achieved by grouping different perturbative terms together 
and relabeling some parameters to include the smoothing-dependent factors,
a procedure which is not uniquely defined.
An altogether different approach which does not need such an operation is the 
Lagrangian resummation theory by \cite{Matsubara:2007wj,Matsubara:2008wx}.

On the other hand, SPT has the advantage of remaining a fully predictive 
theory, where the bias coefficients can be calculated as a function of halo 
mass.
Furthermore, the choice of the smoothing scale $R$ is not completely arbitrary, 
but confined to a rather narrow range around $R \simeq 10\ h^{-1}$~Mpc. 
Indeed, the smoothing needs to be $R \gtrsim 8 \ h^{-1}$~Mpc in order not to break the 
validity of the perturbative expansion in a significant fraction of the volume
($\sigma \ll 1$ for matter and 
$\sigma \ll b_{20}/b_{10}$ for halos) and, on the other hand,
$R$ needs to be as close as possible to this limit if we want to prevent the 
smoothing from wiping out the non-linear corrections at the wavenumbers 
of interest.

We have checked numerically that different choices of the smoothing scale $R$ 
within a reasonable range larger than the Lagrangian size of the halos 
do not affect our results significantly.
Notice that, for non-Gaussian perturbations,
the $k$-independent -- but $R$-dependent --
terms arising in SPT on large scales  are less important, 
since the halo power spectrum grows with decreasing $k$.

\section {Bispectra} \label{sec:bispectrum}

The leading contributions to the halo bispectrum from non-Gaussian initial
conditions of
the local type have
been recently computed in the framework of the local bias model given in
Eq.~(\ref{ldb})
\cite{Sefusatti:2007ih,Sefusatti:2009qh,Jeong:2009vd}:
\be
B_h(\mb k_1, \mb k_2, \mb k_3)=b_1^3
B_\delta(\mb k_1,\mb k_2,\mb k_3)+b_1^2b_2\left[P_\delta(k_1)P_\delta(k_2)+
\mathrm{(2 \ cyc.)}+\frac{1}{2}\int\frac{d^3q}{(2\pi)^3}
T_\delta(\mb q,\mb k_1-\mb q,\mb k_2,\mb k_3) + \mathrm{(2\ cyc.)}\right].
\label{eq:bisploc}
\ee
Using Eulerian perturbation theory to follow the growth of density
perturbations gives up to fourth order
\be
B_\delta(\mb k_1,\mb k_2,\mb k_3)\simeq B_0(\mb k_1,\mb k_2,\mb k_3)+
2 F_2(\mb k_1, \mb k_2) P_0(k_1) P_0(k_2) + \mathrm{ (2\ cyc.)}\;,
\ee
where the second term is generated by non-linear gravity while
\be
B_0(\mb k_1,\mb k_2,\mb k_3)\,\simeq\,
2 \fnl \,    \left[  \alpha (k_3) \, \frac
{P_0(k_1)\,P_0(k_2)}{\alpha (k_1) \, \alpha (k_2)} +  \mathrm{2
\ cyc.} \right] \,
\ee
is the linear matter bispectrum due to primordial non-Gaussianity.
Similarly, the term between square brackets in Eq.~(\ref{eq:bisploc})
reduces to
\be
P_0(k_1) P_0(k_2) + \mathrm{(2\ cyc.)}+\frac{1}{2}\int
\frac{d^3q}{(2\pi)^3} T_0(\mb q,\mb k_1-\mb q,\mb k_2,\mb k_3) +
\mathrm{(2\ cyc.)}\;.
\ee

In full analogy with the power spectrum calculation discussed above,
it is straightforward to show that
our new expansion of $\delta_h$ in terms of both $\delta$ and $\varphi$
gives rise to many additional contributions. The most compact form is
obtained
by writing the expansion of the product of three $\delta_h$ evaluated
in real space at three different locations.
For the term which generates
the contribution to the halo bispectrum which is
proportional to the linear matter bispectrum, we have:
\be
b_1 \delta_1 \delta_1 \delta_1 \to
b_{10} \delta_1 \delta_1 \delta_1 +
b_{10}^2 b_{01}\varphi \delta_1 \delta_1 + \mathrm{(2\ cyc.)} + b_{10}
b_{01}^2 \varphi \varphi \delta + \mathrm{(2\ cyc.)}
+b_{01}^3 \varphi \varphi \varphi\;.
\label{eq:bisp3ord}
\ee
On the other hand, for the term accounting for
the non-linear evolution of the density, one finds:
\be
b_1^3 \delta_1 \delta_1  \delta_2 + \mathrm{(2\ cyc.)} \to
b_{10}^3 \delta_1 \delta_1 \delta_2 + \mathrm{(2\ cyc.)} +
b_{10}^2 b_{01} \varphi\delta_1 \delta_2+ \mathrm{(5\ cyc.)}+ b_{10}
b_{01}^2
\varphi \varphi \delta_2 + \mathrm{(2\ cyc.)}\;.
\label{eq:bisp4orda}
\ee
Finally, for the source of the term between
square brackets in in Eq.~(\ref{eq:bisploc}), we get:
\ba
\frac{1}{2} b_1^2 b_2 \delta_1 \delta_1 \delta_1^2 + \mathrm{(2\ cyc.)}
&\to&
\frac{1}{2}  b_{10}^2 b_{20} \delta_1 \delta_1\delta_1^2+ \mathrm{(2\
cyc.)} +
b_{10}^2 b_{11} \delta_1 \delta_1 (\varphi \delta_1)+ \mathrm{(2\ cyc.)} +
\frac{1}{2} b_{10}b_{20}b_{01} \delta_1\varphi \delta_1^2+ \mathrm{(5\
cyc.)} + \nonumber \\
&&\frac{1}{2}  b_{10}^2 b_{02}\delta_1\delta_1 \varphi^2 + \mathrm{(2\
cyc.)}
+b_{10}b_{01}b_{11}\delta_1\varphi (\varphi \delta_1) + \mathrm{(5\ cyc.)}
 +\frac{1}{2} b_{01}^2 b_{20} \varphi \varphi \delta_1^2+ \mathrm{(2\
cyc.)} +\nonumber \\
&&\frac{1}{2} b_{10}b_{01} b_{02} \delta_1\varphi \varphi^2 +
\mathrm{(5\ cyc.)}+
b_{01}^2b_{11} \varphi\varphi(\varphi\delta_1)+ \mathrm{(2\ cyc.)}+
\frac{1}{2} b_{01}^2 b_{02} \varphi \varphi\varphi^2 + \mathrm{(2\
cyc.)}\;. 
\label{eq:bisp4ordb}
\ea

\begin {figure}
\begin{center}
\includegraphics[width=0.45\columnwidth,angle=0]{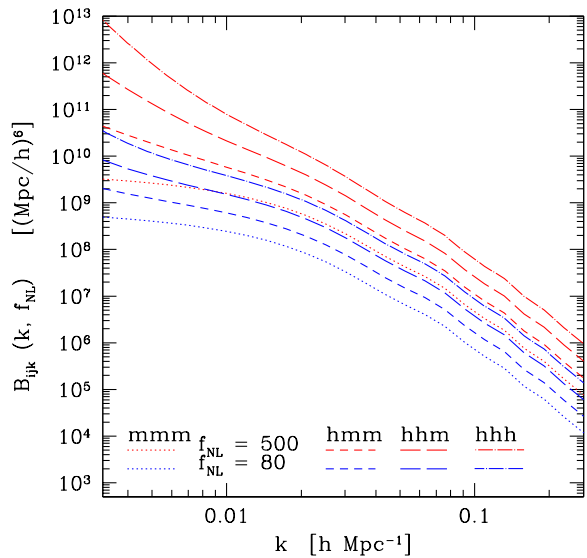}
\end{center}
\caption{Theoretical bispectra at tree level in the equilateral configuration:
$B^{mmm}(k,k,k), B^{hmm}(k,k,k), B^{hhm}(k,k,k)$, $ B^{hhh}(k,k,k)$ for two
values of $\fnl$.  In the Gaussian
case ($\fnl = 0$) the tree-level bispectrum vanishes.}

\label{fig:Bk}
\end{figure}

The halo bispectrum can be computed by Fourier transforming the expressions
above.
For instance, from Eq.~(\ref{eq:bisp3ord}) we obtain:
\be \label{eq:bisp}
B_{h}(\mb k_1,\mb k_2,\mb k_3) \simeq \left[ b_{10} + \frac {b_{01}}
{\alpha(k_1)}  \right]
\left[ b_{10} + \frac {b_{01}} {\alpha(k_2)}  \right] \left[ b_{10} + \frac
     {b_{01}} {\alpha(k_3)}  \right]  B_0(\mb k_1, \mb k_2, \mb k_3) \;,
\ee
while assuming a local-bias scheme would have given $B_h(\mb k_1,\mb k_2,\mb
k_3) \simeq b_{10}^3 \, B_0(\mb k_1, \mb k_2, \mb k_3)$.
The mixed matter-halo bispectra can be obtained in a similar way, and their
expressions differ from Eq.~(\ref{eq:bisp}) only for the presence of a
reduced number of scale-dependent bias factors. We plot the equilateral
configuration of the bispectra in Fig.~\ref{fig:Bk}, for two values of
$\fnl$. Since we are only considering the tree-level contribution,
the Gaussian bispectrum is vanishing.
Note that our leading-order result of Eq.~(\ref{eq:bisp}) can be
reproduced by
taking  the analogous formula obtained with the univariate local bias and
simply replacing $b_{10}$ with the
scale-dependent bias $b_{10}+b_{01}/\alpha(k)$.
More complex equations relate the scale-dependent terms obtained by
taking the
Fourier transform of Eqs. (\ref{eq:bisp4orda}) and (\ref{eq:bisp4ordb}).
We will not discuss them in detail here.

It is important to notice that the scale-dependent bias changes the shape
dependence of the halo bispectrum. The tree-level term $B_0$
is dominated by the squeezed configurations (where one of the wavevectors is
small) and the $\fnl$-dependent term $b_{01}/\alpha$ makes it even more so.
Comparing our results with the figures in \cite{Jeong:2009vd}
suggests that, in strict analogy with the result for the power spectrum,
the terms proportional to $b_{01}/\alpha$ give by far the
dominant contribution to the bispectrum on large scales.
The shapes of the bispectrum parts coming from the non-linear
growth of perturbations and second-order biasing are
very different \cite{Sefusatti:2007ih, Jeong:2009vd} and
this makes the bispectrum a promising tool to measure $\fnl$.

We can see that the expanded form of the bispectrum deriving from Eqs. (\ref{eq:bisp4orda}) and (\ref{eq:bisp4ordb}) depends on all the non-Gaussianity parameters in a non-trivial way: all terms involving the matter bispectrum $B_0$, i.e. all terms involving an average over three $\delta$'s or $\varphi$'s bring in a linear dependence on $\fnl$. Then, all terms involving the matter trispectrum $T_0$, i.e. averages over four $\delta$'s or $\varphi$'s, have two contributions, depending  on $\gnl$ and $\fnl^2$ (or in general $\tnl$) respectively. In addition to this, we have additional dependences on $\fnl$ and $\gnl$ implicit in the bias factors, as described in Section \ref{sec:bias}. 
This complex shape and scale dependence of the bispectrum offers
an unique opportunity to simultaneously constrain
$\fnl$, $\gnl$ and $\tnl$ and thus distinguish between
 inflationary models.
We will investigate this in the near future.

\section{Conclusions} \label {sec:concl}
We have studied the growth of structure from
non-Gaussian initial conditions of the local type.
In particular,
we have shown that the spatial distribution of dark-matter halos is naturally
described by a multivariate local bias scheme where the halo number density 
depends on the underlying values of the density field $\delta$, the auxiliary Gaussian
potential $\varphi$, and (possibly) also on its gradient.
This bivariate local approach can be equally interpreted as a non-local description in terms of $\delta$ only, since $\varphi$ and $\delta$ are related by the Poisson equation.
Adopting the peak-background split, some common parameterizations of 
the halo mass function, and a local model for the evolution of large-scale
density perturbations,
we have derived the coefficients of this multivariate
expansion as a function of the halo mass and of the parameters quantifying
the level of primordial non-Gaussianity.

Using SPT to approximate the non-linear growth of density
perturbations, we have computed the halo power spectrum and the halo-matter
cross spectrum up to the third non-vanishing perturbative order.
For unbiased tracers our result coincides with the matter
power spectrum presented by \cite{Taruya:2008pg}. 
However, in the most general (biased) case, it differs from what is obtained adopting
the standard local bias expansion in terms of the density field
\cite{Taruya:2008pg, Sefusatti:2009qh, Sefusatti:2007ih, Jeong:2009vd}.
The most remarkable feature is that 
the scale-dependent bias first discussed in Dal07 
appears at leading order in our model for the power spectrum.
This is because in our multivariate biasing scheme halo fluctuations on
large scales trace the Gaussian potential $\varphi$ rather than $\delta$.
However, our model
reduces to the usual univariate case on larger scales, where the variance
of density fluctuations is much larger than that of the potential.
Note that both the multivariate and the univariate models predict that
the dominant contribution to the halo power spectrum scales as
$\fnl\, P_0(k)/\alpha(k)$ for $k\to 0$. 
However, for the standard univariate biasing, the amplitude of this term 
is given by a badly behaved integral which strongly depends 
on the assumed smoothing scale. 
Renormalization of the second bias coefficient (which should then be treated as 
a fitting parameter when comparing the theory to observation or simulations) 
is unavoidable in this case, while it is not needed in our model.

We have then tested our results against the N-body simulations by PPH08,
finding excellent agreement for both the matter and the halo two-point functions.
Focusing on the scale-dependent bias generated by primordial non-Gaussianity,
we have shown that our model
accounts for the discrepancies previously found between
the predictions by Dal07 and the outcome of numerical simulations 
\cite {Desjacques:2008vf, Pillepich:2008ka, Grossi:2009an}.
Corrections to the simpler model by Dal07 arise for two main
reasons: (a) the bias coefficient $b_{10}$ depends on $\fnl$ 
due to the fact that the shape of the mass function
is altered by primordial non-Gaussianity (see also
\cite{Slosar:2008hx, Desjacques:2008vf, Desjacques:2009jb}), 
and this adds a scale-independent offset to the bias deviation $\Delta b$;
(b) considering perturbation theory up to third order 
generates numerous additional corrective factors that become important on 
intermediate and small scales.
With our one-loop calculation of the power spectrum, 
the range of validity of the theory extends up to scales 
$k \sim 0.1-0.3\ h$ Mpc$^{-1}$  depending on halo
mass and redshift.

We have also shown how our calculations can be extended to include
higher-order terms of primordial non-Gaussianity,
for instance by considering a non-vanishing primordial trispectrum 
proportional to the parameter $\gnl$.
In this case, 
the halo power spectrum includes an additional contribution 
proportional to $\gnl^2 P_0(k)/\alpha^2(k)$ which, 
depending on the values of $\fnl$ and
$\gnl$, may be dominant on the largest scales.
This term only appears in our multivariate expansion and 
originates from the bias parameter $b_{02}$ which includes
a correction proportional to $g_{\rm NL}$. 
On the other hand, in agreement with \cite{Desjacques:2009jb},
we have found that both the halo-halo and halo-matter spectra acquire 
a dependence on $\gnl$ from the trispectrum of the linear density field. 
This term scales as $\gnl P_0(k)/\alpha(k)$ but its normalization 
depends on the assumed smoothing scale and cannot be robustly predicted
by the theory.

Finally, we 
have calculated the halo bispectrum deriving from our multivariate
biasing scheme. At tree level, our result corresponds to the usual
bispectrum deriving from Gaussian initial conditions but where 
the linear bias $b_{10}$ is replaced by  $b_{10} + b_{01}/\alpha(k)$. 
This is different from what has been found assuming univariate local biasing
\cite{Sefusatti:2007ih, Jeong:2009vd}. Therefore the analysis of
three-point statistics represents a promising tool to test the different
biasing schemes against observations.
Also, the complex shape and scale dependence of the halo bispectrum offers
an unique opportunity to simultaneously constrain
$\fnl$, $\gnl$ and $\tau_{\mathrm NL}$ and thus put entire classes
of inflationary models under scrutiny.
We will explore this in more detail in a forthcoming paper.

\section*{Acknowledgments}
We thank Christian T. Byrnes and Kazuya Koyama for useful discussions and comments on the draft, and Annalisa Pillepich for help with the simulation data.
TG acknowledges support from the Alexander von Humboldt Foundation.

 \bibliography{ms}

\appendix

\section* {Appendix: Complete analytic expression of the power spectra} \label {appendix}
We list below the non-vanishing contributions to the halo-halo and
halo-matter power spectra up to one-loop in perturbation theory. 
These have been obtained by smoothing the evolved density perturbations with the
filter $W(k R)$, so that $\tilde\delta (k) \rightarrow \tilde \delta (k, R)
= \tilde\delta (k) \, W (k R)$. 
We highlight with the label ``Local''
the terms that are also present if we use the univariate local bias
approach as in \cite{Taruya:2008pg}. Similarly, the contributions that
do not vanish in the Gaussian case are marked 
with the label ``Gauss''.  
We have dropped the two-loop terms which arise from the trispectrum $T_0$, 
as they  generally give negligible contributions, with the exception of the term $P_{13}^{II} (k, R)$,
 which is important in the case of large $\gnl$ and small $\fnl$.
Some of the integrals below present an infrared divergence if $n_s \simeq 1$, like for instance the term $P^{hh}_{(23)(23)}$. In this case, we introduce a cutoff in $P_0(k)$ for $k < k_H = 1/R_H$ where $R_H = c/H_0$ \cite{1999RSPTA.357...57B}.

Note that only the halo-matter cross spectrum has been compared to the N-body
results by PPH08, as in the simulations
the halo-halo spectrum is more strongly
affected by shot noise.

\subsection{The halo-halo spectrum}

The full halo-halo power spectrum at one loop is
\be
P^{hh}(k,z,R) =  D^2(z) \, P^{hh}_{11}(k,R) + D^3(z) \, P^{hh}_{12}(k,R) + D^4(z) \, \left[ P^{hh}_{22}(k,R) + P^{hh}_{13}(k,R) \right],
\ee 
where:

$P^{hh}_{11}(k,R)$ is the sum of the following terms:
\ba \label{eq:big1}
 \mbox{Local, Gauss}\:\:\:\:   P^{hh}_{(10)(10)}  (k, R) &=&  b_{10}^2 \, P_0(k) \, W^2 (k R) \nonumber \\
   2 P^{hh}_{(10)(11)} (k, R) &=& 2 \, b_{10} \, b_{01} \, \frac{P_0(k)}{\alpha(k)} \, W^2 (k R) \nonumber \\
   P^{hh}_{(11)(11)} (k, R)   &=&  b_{01}^2 \, \frac{P_0(k)}{\alpha^2(k)} \, W^2 (k R). 
\ea

$P^{hh}_{12}(k,R)$ is the sum of the following terms:
\ba
\mbox{Local}\:\:\:\: 2 P^{hh}_{(10)(20)}(k, R) &=& 2 \, b_{10}^2 \, \int \frac{d^3 \mathbf q}{(2 \pi)^3}  \, B_0 (\mathbf k, \mathbf q, -\mathbf k - \mathbf q )  \, J_2^{(s)}(-\mathbf k - \mathbf q, \mathbf q)  \,  W^2(k R) \, W^2(q R)    \nonumber \\
2 P^{hh}_{(11)(20)}(k, R) &=&   2 \, b_{01} \, b_{10} \, \int \frac{d^3 \mathbf q}{(2 \pi)^3}  \, \frac {B_0 (\mathbf k, \mathbf q, -\mathbf k - \mathbf q )} {\alpha (k)}  \, J_2^{(s)}(-\mathbf k - \mathbf q, \mathbf q)  \,  W^2(k R) \, W^2(q R)    \nonumber \\
\mbox{Local}\:\:\:\: 2 P^{hh}_{(10)(21)}(k, R) &=&   b_{10} \, b_{20}  \, \int \frac{d^3 \mathbf q}{(2 \pi)^3}  \,  B_0 (\mathbf k, \mathbf q, -\mathbf k - \mathbf q )  \,  W(k R) \, W(q R) \, W (|\mathbf{k}+\mathbf{q}| R)  \nonumber \\
2 P^{hh}_{(11)(21)}(k, R) &=&   b_{01} \, b_{20}  \, \int \frac{d^3 \mathbf q}{(2 \pi)^3}  \,  \frac{B_0 (\mathbf k, \mathbf q, -\mathbf k - \mathbf q )}{\alpha(k)}  \,  W(k R) \, W(q R) \, W (|\mathbf{k}+\mathbf{q}| R)  \nonumber \\
2 P^{hh}_{(10)(22)}(k, R) &=&   b_{10} \, b_{11}  \, \int \frac{d^3 \mathbf q}{(2 \pi)^3}  \, \frac {B_0 (\mathbf k, \mathbf q, -\mathbf k - \mathbf q )} {\alpha (q)}  \,  W(k R) \, W(q R) \, W (|\mathbf{k}+\mathbf{q}| R)   \nonumber \\
2 P^{hh}_{(11)(22)}(k, R) &=&   b_{01} \, b_{11} \, \int \frac{d^3 \mathbf q}{(2 \pi)^3}  \, \frac {B_0 (\mathbf k, \mathbf q, -\mathbf k - \mathbf q )} {\alpha(k) \, \alpha (q)}  \,  W(k R) \, W(q R) \, W (|\mathbf{k}+\mathbf{q}| R)   \nonumber \\
2 P^{hh}_{(10)(23)}(k, R) &=&   b_{10} \,  b_{02}  \, \int \frac{d^3 \mathbf q}{(2 \pi)^3}  \, \frac {B_0 (\mathbf k, \mathbf q, -\mathbf k - \mathbf q )} {\alpha (q)  \, \alpha (| \mathbf k + \mathbf q|)}  \,  W(k R) \, W(q R) \, W (|\mathbf{k}+\mathbf{q}| R)  \nonumber \\
2 P^{hh}_{(11)(23)}(k, R) &=&   b_{01} \,  b_{02}  \, \int \frac{d^3 \mathbf q}{(2 \pi)^3}  \, \frac {B_0 (\mathbf k, \mathbf q, -\mathbf k - \mathbf q )} {\alpha(k) \, \alpha (q)  \, \alpha (| \mathbf k + \mathbf q|)}  \,  W(k R) \, W(q R) \, W (|\mathbf{k}+\mathbf{q}| R). 
\ea

$P^{hh}_{22}(k,R)$ is the sum of the following terms:
\ba
\mbox{Local, Gauss}\:\:\:\:    P^{hh}_{(20)(20)} (k, R)   &=&  2 \, b_{10}^2 \, \int \frac{d^3 \mathbf q}{(2 \pi)^3} \, P_0(q) \, P_0(|\mathbf k - \mathbf q|) \, \left[ J_2^{(s)}(\mathbf q, \mathbf k - \mathbf q) \right]^2  \, W^2 (k R) \nonumber \\
\mbox{Local, Gauss}\:\:\:\:    2 P^{hh}_{(20)(21)} (k, R)   &=& 2 \, b_{10} \,  b_{20} \, \int \frac{d^3 \mathbf q}{(2 \pi)^3} \,  P_0(q) \, P_0(|\mathbf k - \mathbf q|) \, J_2^{(s)}(\mathbf q, \mathbf k - \mathbf q) \, W(k R) \, W(q R) \, W (|\mathbf{k}-\mathbf{q}| R)     \nonumber \\
\mbox{Local, Gauss}\:\:\:\:    P^{hh}_{(21)(21)} (k, R)   &=&  \frac 1 2 \, b_{20}^2 \, \int \frac{d^3 \mathbf q}{(2 \pi)^3} \, P_0(q) \, P_0(| \mathbf k - \mathbf q|) \, W^2(q R) \, W^2 (|\mathbf{k}-\mathbf{q}| R)   \nonumber \\
2P^{hh}_{(20)(22)} (k, R) &=& 2 \, b_{10} \, b_{11} \, \int \frac{d^3 \mathbf q}{(2 \pi)^3} \, \frac {P_0(q)}{\alpha (q)} \, P_0(| \mathbf k - \mathbf q|) \, J_2^{(s)}(\mathbf q, \mathbf k - \mathbf q)  \, W(k R) \, W(q R) \, W (|\mathbf{k}-\mathbf{q}| R)   \nonumber \\
2P^{hh}_{(21)(22)} (k, R) &=&  b_{20} \, b_{11} \, \int \frac{d^3 \mathbf q}{(2 \pi)^3} \,  \frac{P_0(q)}{\alpha(q)} \, P_0(|\mathbf k - \mathbf q|)  \, W^2(q R) \, W^2 (|\mathbf{k}-\mathbf{q}| R)   \nonumber \\
P^{hh}_{(22)(22)} (k, R) &=&  \frac 1 4 \, b_{11}^2  \int \frac{d^3 \mathbf q}{(2 \pi)^3} \, \left[ P_0(q) \, \frac{P_0(| \mathbf k - \mathbf q|)}{\alpha(|\mathbf k - \mathbf q|)^2} + \frac{P_0(q)}{\alpha(q)} \, \frac{P_0(| \mathbf k - \mathbf q|)}{\alpha(|\mathbf k - \mathbf q|)} \right]  \, W^2(q R) \, W^2 (|\mathbf{k}-\mathbf{q}| R)  \nonumber \\
2P^{hh}_{(20)(23)} (k, R) &=& 2 \, b_{10} \, b_{02} \,  \int \frac{d^3 \mathbf q}{(2 \pi)^3} \, \frac{P_0(q)}{\alpha(q)} \, \frac{P_0(|\mathbf k - \mathbf q|)}{\alpha(| \mathbf k - \mathbf q|)} \, J_2^{(s)}( \mathbf q, \mathbf k - \mathbf q) \, W(k R) \, W(q R) \, W (|\mathbf{k}-\mathbf{q}| R)   \nonumber \\
2P^{hh}_{(21)(23)} (k, R) &=&  b_{20} \, b_{02} \,  \int \frac{d^3 \mathbf q}{(2 \pi)^3} \, \frac{P_0(q)}{\alpha(q)} \,  \frac{P_0(|\mathbf k - \mathbf q|)}{\alpha(|\mathbf k - \mathbf q|)} \, W^2(q R) \, W^2 (|\mathbf{k}-\mathbf{q}| R) \nonumber \\
2P^{hh}_{(22)(23)} (k, R) &=&    b_{11} \, b_{02} \, \int \frac{d^3 \mathbf q}{(2 \pi)^3} \,  \frac{P_0(q)}{\alpha(q)} \, \frac{P_0(| \mathbf k - \mathbf q|)}{\alpha^2(| \mathbf k - \mathbf q|) } \, W^2(q R) \, W^2 (|\mathbf{k}-\mathbf{q}| R) \nonumber \\
P^{hh}_{(23)(23)}(k, R) &=& \frac 1 2 \, b_{02}^2  \, \int \frac{d^3 \mathbf q}{(2 \pi)^3} \, \frac{P_0(q)}{\alpha^2(q)} \frac{P_0(|\mathbf k - \mathbf q|)}{\alpha^2(|\mathbf k - \mathbf q|) } \, W^2(q R) \, W^2 (|\mathbf{k}-\mathbf{q}| R). 
\ea
$P^{hh}_{13}(k,R)$ is the sum of the following terms:
\ba
\mbox{Local, Gauss}\:\:\:\:   2 P^{hh}_{(10)(30)} (k, R) &=& 6 \, b_{10}^2  \,  P_0(k) \, \int \frac{d^3 \mathbf q}{(2 \pi)^3} \, P_0(q) \, J_3^{(s)}(\mathbf k, \mathbf q, -\mathbf q) \, W^2 (k R)   \nonumber \\
2 P^{hh}_{(11)(30)} (k, R) &=& 6 \, b_{01} \, b_{10} \, \frac{P_0(k)}{\alpha(k)} \,  \int \frac{d^3 \mathbf q}{(2 \pi)^3} \, P_0(q) \, J_3^{(s)}(\mathbf k, \mathbf q, -\mathbf q) \,  W^2 (k R)  \nonumber \\
\mbox{Local, Gauss}\:\:\:\:    2 P^{hh}_{(10)(31)} (k, R) &=& 4 \, b_{10} \, b_{20} \, P_0(k) \,  \int \frac{d^3 \mathbf q}{(2 \pi)^3} \, P_0(q) \,  J_2^{(s)}(\mathbf q, \mathbf k) \, W(k R) \, W(q R) \, W (|\mathbf{k} + \mathbf{q}| R)  \nonumber \\
2 P^{hh}_{(11)(31)} (k, R) &=& 4 \, b_{01} \, b_{20}  \,\frac{P_0(k)}{\alpha(k)} \,  \int \frac{d^3 \mathbf q}{(2 \pi)^3} \,  P_0(q) \, J_2^{(s)}(\mathbf q, \mathbf k)  \, W(k R) \, W(q R) \, W (|\mathbf{k} + \mathbf{q}| R)   \nonumber \\
2 P^{hh}_{(10)(32)} (k, R) &=& 2 \, b_{10} \, b_{11} \,  P_0(k) \, \int \frac{d^3 \mathbf q}{(2 \pi)^3} \,  \frac{P_0(q)}{\alpha(q)} \,  J_2^{(s)}(\mathbf q, \mathbf k) \, W(k R) \, W(q R) \, W (|\mathbf{k} + \mathbf{q}| R)  \nonumber \\
2 P^{hh}_{(11)(32)} (k, R) &=& 2 \, b_{01} \, b_{11} \, \frac {P_0(k)}{\alpha (k)} \, \int \frac{d^3 \mathbf q}{(2 \pi)^3} \,  \frac{P_0(q)}{\alpha(q)} \,  J_2^{(s)}(\mathbf q, \mathbf k)  \, W(k R) \, W(q R) \, W (|\mathbf{k} + \mathbf{q}| R)  \nonumber  \\
2 P^{hh}_{(10)(33)} (k, R) &=& \frac 2 3 \, b_{10} \, b_{21} \,  P_0(k) \int \frac{d^3 \mathbf q}{(2 \pi)^3} \, \frac {P_0(q)} {\alpha(q)} \, W^2 (k R) \, W^2 (q R)  \nonumber \\
2 P^{hh}_{(11)(33)} (k, R) &=& \frac 2 3 \, b_{01} \, b_{21} \,  \frac {P_0(k)}{\alpha(k)} \int \frac{d^3 \mathbf q}{(2 \pi)^3} \, \frac {P_0(q)} {\alpha(q)} \, W^2 (k R) \, W^2 (q R)  \nonumber \\
2 P^{hh}_{(10)(34)} (k, R) &=&  \frac 1 3 \, b_{10} \, b_{12} \, P_0(k) \, \int \frac{d^3 \mathbf q}{(2 \pi)^3} \, P_0(q)  \, \left[ \frac {1} {\alpha^2(q)} \, + \,  \frac {1}{\alpha(k) \, \alpha(q)} \right] \, W^2 (k R) \, W^2 (q R)  \nonumber \\
2 P^{hh}_{(11)(34)} (k, R) &=&   \frac 1 3 \, b_{01} \, b_{12} \, \frac{P_0(k)}{\alpha(k)} \, \int \frac{d^3 \mathbf q}{(2 \pi)^3} \, P_0(q) \, \left[ \frac {1} {\alpha^2(q)} \, + \, \frac {1}{\alpha(k) \, \alpha(q)} \right] \, W^2 (k R) \, W^2 (q R)  \nonumber 
\ea
\ba
\mbox{Local, Gauss}\:\:\:\:   2 P^{hh}_{(10)(35)} (k, R) &=&   b_{10} \, b_{30} \, P_0(k) \, \int \frac{d^3 \mathbf q}{(2 \pi)^3} \, P_0(q) \, W^2 (k R) \, W^2 (q R)  \nonumber \\
2 P^{hh}_{(11)(35)} (k, R) &=&  b_{01} \, b_{30} \, \frac {P_0(k)}{\alpha(k)} \, \int \frac{d^3 \mathbf q}{(2 \pi)^3} \, P_0(q) \, W^2 (k R) \, W^2 (q R)  \nonumber \\
2 P^{hh}_{(10)(36)}(k, R) &=&  b_{10} \, b_{03} \, \frac {P_0(k)}{\alpha(k)} \, \int \frac{d^3 \mathbf q}{(2 \pi)^3} \, \frac {P_0(q)}{\alpha^2(q)} \, W^2 (k R) \, W^2 (q R)   \nonumber \\
2 P^{hh}_{(11)(36)}(k, R) &=&  b_{01} \, b_{03} \, \frac {P_0(k)}{\alpha^2(k)} \, \int \frac{d^3 \mathbf q}{(2 \pi)^3} \, \frac {P_0(q)}{\alpha^2(q)} \, W^2 (k R) \, W^2 (q R).  
\ea

The only important two-loop contribution is (relevant for small $\fnl$, large $\gnl$):
\be \label {eq:PII13}
P_{13}^{hh, II} (k, R) = \frac 1 3 \, b_{10} b_{30} \int \frac {d^3 \mb q \, d^3 \mb p} {(2 \pi)^6} \, T_0 (\mb k, \mb q, \mb p, \mb{-k -q -p}) \, W (q R) \, W(p R) \, W (|\mb q + \mb p| R) \, W(|-\mb k - \mb q - \mb p| R).
\ee

\subsection{The halo-matter spectrum}

The full halo-matter cross spectrum at one loop is
\be
P^{hm}(k,z,R) =  D^2(z) \, P^{hm}_{11}(k,R) + D^3(z) \, P^{hm}_{12}(k,R) + D^4(z) \, \left[ P^{hm}_{22}(k,R) + P^{hm}_{13}(k,R) \right].
\ee 
$P^{hm}_{11}(k,R)$ is the sum of the following terms:
\ba \label{eq:big1hm}
 \mbox{Local, Gauss}\:\:\:\:   P^{hm}_{(10)(10)}  (k, R) &=&  b_{10} \, P_0(k) \, W^2 (k R) \nonumber \\
     P^{hm}_{(10)(11)} (k, R) &=&     b_{01} \, \frac{P_0(k)}{\alpha(k)} \, W^2 (k R). 
\ea

$P^{hm}_{12}(k,R)$ is the sum of the following terms:
\ba
\mbox{Local}\:\:\:\: 2 P^{hm}_{(10)(20)}(k, R) &=& 2 \, b_{10}  \, \int \frac{d^3 \mathbf q}{(2 \pi)^3}  \, B_0 (\mathbf k, \mathbf q, -\mathbf k - \mathbf q )  \, J_2^{(s)}(-\mathbf k - \mathbf q, \mathbf q)  \,  W^2(k R) \, W^2(q R)    \nonumber \\
P^{hm}_{(11)(20)}(k, R) &=&   b_{01}  \, \int \frac{d^3 \mathbf q}{(2 \pi)^3}  \, \frac {B_0 (\mathbf k, \mathbf q, -\mathbf k - \mathbf q )} {\alpha (k)}  \, J_2^{(s)}(-\mathbf k - \mathbf q, \mathbf q)  \,  W^2(k R) \, W^2(q R)    \nonumber \\
\mbox{Local}\:\:\:\: P^{hm}_{(10)(21)}(k, R) &=&  \frac 1 2 \, b_{20}  \, \int \frac{d^3 \mathbf q}{(2 \pi)^3}  \,  B_0 (\mathbf k, \mathbf q, -\mathbf k - \mathbf q )  \,  W(k R) \, W(q R) \, W (|\mathbf{k}+\mathbf{q}| R)  \nonumber \\
P^{hm}_{(10)(22)}(k, R) &=&  \frac 1 2 \, b_{11}  \, \int \frac{d^3 \mathbf q}{(2 \pi)^3}  \, \frac {B_0 (\mathbf k, \mathbf q, -\mathbf k - \mathbf q )} {\alpha (q)}  \,  W(k R) \, W(q R) \, W (|\mathbf{k}+\mathbf{q}| R)   \nonumber \\
P^{hm}_{(10)(23)}(k, R) &=&  \frac 1 2 \, b_{02}  \, \int \frac{d^3 \mathbf q}{(2 \pi)^3}  \, \frac {B_0 (\mathbf k, \mathbf q, -\mathbf k - \mathbf q )} {\alpha (q)  \, \alpha (| \mathbf k + \mathbf q|)}  \,  W(k R) \, W(q R) \, W (|\mathbf{k}+\mathbf{q}| R). 
\ea

$P^{hm}_{22}(k,R)$ is the sum of the following terms:
\ba
\mbox{Local, Gauss}\:\:\:\:    P^{hm}_{(20)(20)} (k, R)   &=&  2 \, b_{10} \,  \int \frac{d^3 \mathbf q}{(2 \pi)^3} P_0(q) \, P_0(|\mathbf k - \mathbf q|) \, \left[ J_2^{(s)}(\mathbf q, \mathbf k - \mathbf q) \right]^2   \, W^2 (k R) \nonumber \\
\mbox{Local, Gauss}\:\:\:\:     P^{hm}_{(20)(21)} (k, R)   &=&   b_{20} \, \int \frac{d^3 \mathbf q}{(2 \pi)^3}  \, P_0(q)  \, P_0(|\mathbf k - \mathbf q|) \, J_2^{(s)}(\mathbf q, \mathbf k - \mathbf q) \, W(k R) \, W(q R) \, W (|\mathbf{k}-\mathbf{q}| R)  \nonumber \\
 P^{hm}_{(20)(22)} (k, R) &=&  b_{11}  \, \int \frac{d^3 \mathbf q}{(2 \pi)^3} \, \frac{P_0(q)}{\alpha(q)} \, P_0(| \mathbf k - \mathbf q|)  \, J_2^{(s)}(\mathbf q, \mathbf k - \mathbf q)  \, W(k R) \, W(q R) \, W (|\mathbf{k}-\mathbf{q}| R)   \nonumber \\
P^{hm}_{(20)(23)} (k, R) &=&   b_{02}  \,   \int \frac{d^3 \mathbf q}{(2 \pi)^3} \,  \frac{P_0(q)}{\alpha(q)}  \,  \frac{P_0(|\mathbf k - \mathbf q|)}{\alpha(| \mathbf k - \mathbf q|)}  \,  J_2^{(s)}( \mathbf q, \mathbf k - \mathbf q)  \,  W(k R) \, W(q R) \, W (|\mathbf{k}-\mathbf{q}| R). 
\ea

$P^{hm}_{13}(k,R)$ is the sum of the following terms:
\ba
\mbox{Local, Gauss}\:\:\:\:   2  P^{hm}_{(10)(30)} (k, R) &=& 6 \, b_{10} \, P_0(k)  \, \int \frac{d^3 \mathbf q}{(2 \pi)^3}  \, P_0(q)  \, J_3^{(s)}(\mathbf k, \mathbf q, -\mathbf q)  \, W^2(k R)   \nonumber \\
P^{hm}_{(11)(30)} (k, R) &=& 3  \, b_{01}  \, \frac{P_0(k)}{\alpha(k)}  \,  \int \frac{d^3 \mathbf q}{(2 \pi)^3}  \, P_0(q)  \, J_3^{(s)}(\mathbf k, \mathbf q, -\mathbf q)  \,  W^2(k R) \nonumber \\
\mbox{Local, Gauss}\:\:\:\:  P^{hm}_{(10)(31)} (k, R) &=& 2 \, b_{20} \, P(k) \,  \int \frac{d^3 \mathbf q}{(2 \pi)^3}  \,  P_0(q) \,  J_2^{(s)}(\mathbf q, \mathbf k) \, W(k R) \, W(q R) \, W (|\mathbf{k} + \mathbf{q}| R)  \nonumber \\
P^{hm}_{(10)(32)} (k, R) &=&  b_{11} \,  P_0(k) \int \frac{d^3 \mathbf q}{(2 \pi)^3}  \,  \frac{P_0(q)}{\alpha(q)} \,  J_2^{(s)}(\mathbf q, \mathbf k) \,  W(k R) \, W(q R) \, W (|\mathbf{k} + \mathbf{q}| R)   \nonumber \\
 P^{hm}_{(10)(33)} (k, R) &=& \frac 1 3 \, b_{21} \, P_0(k) \, \int \frac{d^3 \mathbf q}{(2 \pi)^3} \frac {P_0(q)} {\alpha(q)} \,  W^2(k R) \, W^2(q R) \nonumber \\
 P^{hm}_{(10)(34)} (k, R) &=&  \frac 1 6 \, b_{12} \,  P_0(k) \, \int \frac{d^3 \mathbf q}{(2 \pi)^3} \,  P_0(q) \, \left[ \frac {1} {\alpha^2(q)} +  \frac {1}{\alpha(k) \, \alpha(q)} \right] \,  W^2(k R) \, W^2(q R)  \nonumber \\
\mbox{Local, Gauss}\:\:\:\:    P^{hm}_{(10)(35)} (k, R) &=&  \frac 1 3 \, b_{30} \, P_0(k) \, \int \frac{d^3 \mathbf q}{(2 \pi)^3}  \, P_0(q)  \,  W^2(k R) \, W^2(q R)  \nonumber \\
P^{hm}_{(10)(36)}(k, R) &=& \frac 1 3 \, b_{03} \, \frac {P_0(k)}{\alpha(k)} \, \int \frac{d^3 \mathbf q}{(2 \pi)^3} \, \frac {P_0(q)}{\alpha^2(q)}   \,  W^2(k R) \, W^2(q R).  
\ea
The only important two-loop contribution is (relevant for small $\fnl$, large $\gnl$):
\be \label{eq:PII13b}
P_{13}^{hm, II} (k, R) = \frac 1 6 \, b_{30} \int \frac {d^3 \mb q \, d^3 \mb p} {(2 \pi)^6} \, T_0 (\mb k, \mb q, \mb p, \mb{-k -q -p}) \, W (q R) \, W(p R) \, W (|\mb q + \mb p| R) \, W(|-\mb k - \mb q - \mb p| R).
\ee

\label{lastpage}

\end{document}